\documentclass{article}
\usepackage{timet,color}
\usepackage[urlcolor=blue,citecolor=black,linkcolor=black]{hyperref}
\usepackage{natbib}
\usepackage{amsmath,amssymb,amsfonts}
\usepackage[]{graphicx}
\usepackage{authblk}
\usepackage[margin=25mm]{geometry}
\newcommand*{\reviewerA}[1]{\textcolor{black}{#1}}
\newcommand*{\reviewerB}[1]{\textcolor{black}{#1}}
\providecommand{\keywords}[1]{\textbf{\textit{Index terms---}} #1}

\title{Seismic Wavefield Reconstruction based on Compressed Sensing using Data-Driven Reduced-Order Model}
\author[T. Nagata et al.]
  {T. Nagata$^1$, K. Nakai$^1$, K. Yamada$^1$, Y. Saito$^1$, T. Nonomura$^1$,\\ M. Kano$^2$, S. Ito$^3$, and H. Nagao$^3$ \\
  $^1$ Graduate School of Engineering, Tohoku University, 6-6-01, Aramaki-aza-aoba, Aoba-ku, \\Sendai, Miyagi 980-8579, Japan \\
  $^3$ Graduate School of Science, Tohoku University, 6-3, Aramaki-aza-aoba, Aoba-ku, \\Sendai, Miyagi, 980-8578, Japan \\
  $^3$ Earthquake Research Institute, The University of Tokyo, 1-1-1, Yayoi, Bunkyo-ku, Tokyo 113-0032, Japan  
  }

\begin{document}

\label{firstpage}

\maketitle

\renewcommand{\thefootnote}{}
\footnote[0]{This work has been submitted to the Geophysical Journal International for possible publication. Copyright may be transferred without notice, after which this version may no longer be accessible.}

\begin{abstract}
Reconstruction of the distribution of ground motion due to an earthquake is one of the key technologies for the prediction of seismic damage to infrastructure. Particularly, the immediate reconstruction of the spatially continuous wavefield is valuable for decision-making of disaster response decisions in the initial phase. For a fast and accurate reconstruction, utilization of prior information is essential. In fluid mechanics, full-state recovery, which recovers the full state from sparse observation using a data-driven model reduced-order model, is actively used. In the present study, the framework developed in the field of fluid mechanics is applied to seismic wavefield reconstruction.
A seismic wavefield reconstruction framework based on compressed sensing using the data-driven reduced-order model (ROM) is proposed and its characteristics are investigated through numerical experiments. The data-driven ROM is generated from the dataset of the wavefield using the singular value decomposition. The spatially continuous seismic wavefield is reconstructed from the sparse and discrete observation and the data-driven ROM. The observation sites used for reconstruction are effectively selected by the sensor optimization method for linear inverse problems based on a greedy algorithm.
The proposed framework was applied to simulation data of theoretical waveform with the subsurface structure of the horizontally-stratified three layers. The validity of the proposed method was confirmed by the reconstruction based on the noise-free observation. Since the ROM of the wavefield is used as prior information, the reconstruction error is reduced to an approximately lower error bound of the present framework, even though the number of sensors used for reconstruction is limited and randomly selected. In addition, the reconstruction error obtained by the proposed framework is much smaller than that obtained by the Gaussian process regression.
For the numerical experiment with noise-contaminated observation, the reconstructed wavefield is degraded due to the observation noise, but the reconstruction error obtained by the present framework with all available observation sites is close to a lower error bound, even though the reconstructed wavefield using the Gaussian process regression is fully collapsed. Although the reconstruction error is larger than that obtained using all observation sites, the number of observation sites used for reconstruction can be reduced while minimizing the deterioration and scatter of the reconstructed data by combining it with the sensor optimization method.
Hence, a better and more stable reconstruction of the wavefield than randomly selected observation sites can be realized, even if the reconstruction is carried out with a smaller number of observations with observation noise, by combining it with the sensor optimization method.
\end{abstract}


\section{Introduction}
\reviewerB{Prediction of seismic damage to infrastructure is one of the critical issues in earthquake disaster countermeasures. Reconstruction of the distributions of ground motion caused by earthquakes is an effective way for seismic damage prediction. Such technologies allow us to evaluate the robustness of a city and seismic hazards.} In particular, rapid prediction of seismic damage to infrastructure and buildings based on numerical simulation \citep[e.g.,][]{fujita2014quick} is effective for rapid rescue and recovery. Software for estimating seismic damage is developed around the world for the same purpose. For example, the United States Geological Survey developed ShakeMaps, which provides maps of the intensities and ground motion caused by large earthquakes that can be quickly accessible \citep[e.g.,][]{wald1999trinet,wald2005shakemap}. In Taiwan, the Taiwan Earthquake Loss Estimation System software was developed for rapid reconstruction of seismic intensity, infrastructure damage, and casualties \citep[e.g.,][]{yeh2006overview}.

Infrastructure damage is sometimes evaluated by analysing the seismic responses of structures caused by ground motion. This type of analysis requires seismic waveforms at the base of the target structures as boundary conditions, for use in numerically computing the motion of structures. Because direct observations of the ground motions under all structures are not realistic, spatially uniform waveforms based on past earthquakes are often assumed in practical computations. This assumption is obviously invalid for the realistic evaluation of seismic disasters, so a method for estimating the input waveforms for every structure from seismograms at nearby observatories is required. Therefore, our problem can be formulated as seismic wavefield reconstruction, which is a reconstruction of seismic wavefields utilizing an array of seismometers that are distributed more sparsely than the structures.

Various physics-based approaches have been proposed for seismic wavefield imaging \citep[e.g.,][]{boore1972finite,aoi19993d,pitarka19993d,hisada2003theoretical,koketsu2004finite,ichimura2007earthquake}. These approaches solve the wave equation numerically using hypocentre information and subsurface structure models. Although these approaches are powerful for accurate wavefield reconstruction, larger computational cost is required to estimate the high-fidelity wavefield, or a larger model error must be tolerated to estimate the wavefield in a short time. In addition, the immediate estimation is still challenging even if a simple model is used for the simulation.

A data-driven approach is another choice. Interpolation of the observed data is a method for wavefield reconstruction when physics-based approaches are difficult to execute due to the limitation of computational resources. Although it requires assumptions and constraints, traditional methods have achieved some success \citep{kawakami1989simulation,vanmarcke1991conditioned,kameda1994conditioned,kameda1992interpolating,sato1999real}. The restrictions on reconstruction have been greatly reduced in the modern method of the seismic-wave gradiometry (SWG) method \citep[e.g.,][]{langston2007spatial,langston2007wave,liang2009wave,maeda2016reconstruction,shiina2021optimum}. This method is based on the idea that the amplitudes of seismic waves and their spatial gradients at an arbitrary point can be interpolated from the observed amplitudes at surrounding stations without making assumptions concerning velocity structures and locations of earthquakes. The reconstruction method using the ''basis'' that fits the nature of the wavefield is also developed. \citet{sheldrake2002regional} proposes an algorithm for reconstructing teleseismic wave based on wavefield expansion by cubic B-splines. \citet{wilson2007teleseismic} and \citet{zhan2018application} exploited a plane-wave basis to compensate for a lack of spatial resolution of the observation and successfully interpolated the wavefield. \citet{muir2021seismic} proposed a split processing scheme based on a wavelet transform in time and preconditioned curvelet-based compressed sensing in space. Their scheme can obtain a sparse representation of the continuous seismic wavefield. \citet{chen2019obtaining} emploied an iterative rank-reduction method and simultaneously reconstructed the missing part of data. Their method exploits the spatial coherency of three-dimensional data and reconstructs the missing part using the principal components of incomplete data.

The combined method of physics-based and data-driven approaches has also been developed. \citet{kano2017seismic,kano2017seismicA} proposed an seismic wavefield reconstruction framework combining the replica-exchange Monte Carlo method \citep{swendsen1986replica,geyer1991markov,hukushima1996exchange,earl2005parallel} and physics-based seismic simulation with a 1-D subsurface structure \citep{hisada1995efficient}. In their framework, the replica-exchange Monte Carlo method first estimates a posterior distribution of the model parameters related to the subsurface structure and the source, and then the seismic wavefield is reconstructed through a simulation using a set of parameters that maximuzes the posterior. Although their method can accurately reconstruct seismic wavefields in the frequency range less than 1~Hz, it requires repetitive seismic simulations with many different conditions for parameter estimation. Hence, it is still not available for real-time estimation.

For further rapid estimation (i.e., real-time estimation) of the seismic wavefield, the low-cost continuous seismic wavefield reconstruction method with moderate accuracy from discrete observation is valuable. In the field of fluid dynamics, the framework for full state reconstruction by discrete observation and a tailored basis was developed \citep{manohar2018data} as an extension of the discrete empirical interpolation method (DEIM) \citep{chaturantabut2010nonlinear} or \reviewerB{DEIM method based on the QR decomposition of linear algebra} \citep{drmac2016new} in the framework of the Galerkin projection \citep{rowley2004model} that is used for reduced-order modeling. This framework is based on the data-driven reduced-order model (ROM), such as a model constructed using singular-value decomposition (SVD) of the time-series full state. A tailored basis that represents the features of the target extracted from the training dataset is adopted and plays a major role in accurate reconstruction with fewer observations. The reconstruction of spatially continuous data is carried out using a tailored basis and estimated those coefficients. This is so-called ''full state recovery.'' This framework has already been used in the real (experimental) data in the fluid dynamics \citep{kanda2021feasibility,kanda2022proof,zhou2021data,loiseau2018sparse,inoue2021data,jiang2022online,inoba2022optimization,tiwari2022simultaneous,inoue2022data}. Usage of this framework is not limited to the time series data. \citet{kaneko2021data} applied this framework for estimation of the spatial distribution of the propagation time and attenuation rate which are the essential calibration parameters for sound source identification measurements. In their framework, the data-driven ROMs of the distributions of the propagation time and attenuation rate are constructed by many simulations with an expected range of physical parameters. Then, the distributions of the propagation time and attenuation rate are estimated with an arbitrary physical parameter set in the expected range based on ROM and the value of the propagation time and attenuation rate at limited locations.

In addition, sensor optimisation techniques are another important topic for this framework. Various methods have been proposed, and there are methods based on the greedy algorithm \citep{manohar2018data,manohar2018optimal,manohar2019optimized,clark2018greedy,clark2020multi,clark2020sensor,jiang2019group,saito2020data,saito2021determinant,saito2021data,yamada2021fast,yamada2022greedy,nakai2021effect,carter2021data,inoue2021data,li2021efficient,li2021data,nakai2022nondominated,nagata2022randomized,inoue2022data}, the convex relaxation \citep{joshi2009sensor,liu2016sensor,nonomura2021randomized}, and the proximal optimization \citep{fardad2011sparsity,lin2013design,dhingra2014admm,zare2018optimal,nagata2021data,nagata2022data}. 
\reviewerA{The greedy method selects sensor locations by minimizing/maximizing the objective function in each single-sensor subproblem. The convex relaxation method determines the locations of sensors based on the convex and differentiable objective functions. A Boolean-convex problem with nonconvex constraints, which is the original problem, is relaxed to a convex problem with continuous function and convex constraints. The selected sensor location is indicated by the solution vector, which corresponds to weights for sensor candidates. The sensor selection method based on the proximal optimization determines the sensor location by optimizing the objective function including the sparsity promoting term 
.}

In Japan, more than 2000 seismological observatories are in operation and \reviewerB{have been} monitoring and recording the waveform is continuous for over 20 years. In recent years, data obtained from vibrometers installed in buildings and utilities such as facilities for electricity, gas, and water, as well as accelerometers built into smartphones are beginning to be recognized as potential resources for next-generation seismic research \reviewerB{\citep{allen2020myshake}}. In that case, the available observation sites will explosively increase, and it is unreasonable to use all data obtained by all available observation sites. Prior to the advent of the earthquake super-big data era, it is essential to design an observation point selection algorithm that selects data to be used according to the purpose. The importance of sensor selection/placement is beginning to be recognized in the context of seismology. \citet{steinberg2003optimal} consider the network configurations that maximise the precision of the source localisation based on the statistical theory of the optimal design of experiment \citep{atkinson2007optimum}.
\citet{hardt1994design} developed a method to design optimum networks for aftershock recordings based on the technique of simulated annealing \citep{kirkpatrick1983optimization}. This method determined the optimal network configuration that minimises the error of the linearised earthquake location problem based on the D-optimality criterion and was extended by \citet{kraft2013optimization}. \citet{muir2022wavefield} investigated an optimal design strategy for improving the frequency of mixed distributed acoustic sensing network and point sensor deployments for wavefield reconstruction. They solved the problem of deciding the best locations for point sensors given a fixed distributed acoustic sensing network and optimised the network sensitivity by promoting the incoherence of measurements.

The present study attempts to establish the fast and accurate seismic wavefield reconstruction framework that estimates a spatially continuous wavefield from discrete observations. Furthermore, wavefield reconstructions with sparse observations, i.e., observation with effectively-selected observatories, are also conducted. Our framework was inspired by the study of \citet{kaneko2021data} in the field of aeroacoustics. The framework of full state recovery from sparse observation based on the data-driven ROM was extended to the problem of seismic wavefield reconstruction. The spatially continuous and time series seismic wavefield is directly reconstructed from the sparse observations using the ROM. The sensor optimisation method based on the greedy algorithm is used to determine the suitable subset of the available observation sites for reconstruction. 
Sections~\ref{sec:method} and \ref{sec:experiment} respectively describe the data-driven reduced order modelling and full state recovery from the sparse observation based on the data-driven ROM and problem settings of the present numerical experiment. Section~\ref{sec:results} shows the results of the experiments and discusses the characteristics of the proposed method by comparing the obtained results with the reference data and results obtained by the Gaussian process regression. The reconstructed wavefield, reconstruction error, time-series waveform, and frequency spectra are provided. Section~\ref{sec:conclusion} provides concluding remarks.

\section{Methodologies}\label{sec:method}
The present study proposes a seismic wavefield reconstruction framework based on a data-driven ROM and discrete observations. The ROM of the wavefield for a certain region is constructed based on the spatially continuous wavefield data generated by seismic simulations. Many simulations with different parameter sets are conducted, and a data-driven ROM is constructed by means of modal decomposition. In the wavefield reconstruction, the waveform is observed at a limited number of discretely installed sensors, and a continuous wavefield is reconstructed based on the observed waveform and the prepared ROM.

\subsection{Data-driven reduced-order modelling}
In the present study, a data-driven ROM is constructed using the singular value decomposition. Using the singular value decomposition, a matrix decomposition can be obtained as follows:
\begin{align}
\mathbf{X}  &=\mathbf{USV}^{\mathsf T} \nonumber \\
            &\approx \mathbf{U}_r\mathbf{S}_r\mathbf{V}^{\mathsf T}_r
\label{eq:ROM}
\end{align}
where $\mathbf{X}\in\mathbb{R}^{n\times m}$ is the data matrix, \reviewerB{which consists of Fourier spectra of the wavefields} $\mathbf{x}_i\in\mathbb{R}^{n}$ with a spatial dimension of $n$ obtained by simulations with $m$ different parameter sets (i.e., $\mathbf{X}=\left[\mathbf{x}_1\,\,\mathbf{x}_2\,\cdots\, \mathbf{x}_m\right]$). Here, $\mathbf{x}_i$ is vectorized observation data in which observation data are arranged vertically. \reviewerB{The detailed descriptions of the data matrix in the present study appear in Section~\ref{sec:dataset}.}
The data matrix $\mathbf{X}$ is decomposed into a left singular matrix $\mathbf{U}\in\mathbb{R}^{n\times m}$ which shows the spatial modes; a diagonal matrix of singular values $\mathbf{S}\in\mathbb{R}^{m\times m}$ showing the amplitude of the modes; and a right singular matrix $\mathbf{V}\in\mathbb{R}^{m\times m}$.  The subscript $\circ_r$ indicates that the higher modes than the $r$th mode are truncated. Dimensional reduction is conducted by the truncated SVD \citep{eckart1936approximation}, and the rank-$r$ reduced-order modelling can be conducted by the rank-$r$ approximation of the data matrix. The ROM is constructed by truncating higher-order modes, except for the leading-$r$ singular values and vectors. \reviewerA{The rank of ROM is connected to the model error which is the lower bound of the reconstruction error.}

\subsection{Full-state recovery based on discrete observation} \label{sec:full-data_recovery}
A scalar snapshot measurement $\mathbf{x}_p\in\mathbb{R}^p$ through discrete observation of the full state $\mathbf{x}\in\mathbb{R}^n$ including uniform independent Gaussian noise $\mathbf{v}\sim\mathcal{N}(\mathbf{0},\sigma^2\mathbf{I}) \in\mathbb{R}^p$ can be expressed as follows:

\begin{align}
    \mathbf{x}_p=\mathbf{H}\mathbf{x+v},
    \label{eq:observation_org}
\end{align}
where $\sigma^2$, $\mathbf{x}_p$, and $\mathbf{H}\in\mathbb{R}^{p\times n}$ are the noise variance, the observation vector, and the sensor location matrix, respectively. Here, $p$ and $n$ are the number of selected sensors and the number of potential sensor locations, respectively. The selected sensors indicate the observation sites used for the full state recovery, and the potential sensors indicate all available observation sites. The sensor location matrix $\mathbf{H}$ is a sparse matrix that indicates selected sensor locations. Each row vector of $\mathbf{H}$ is a unit vector and the locations of the unity in the row vectors correspond to the locations which are chosen from $n$ potential sensor locations. \reviewerB{In the present study, the sensor location matrix $\mathbf{H}$ is constructed by random selection or the greedy-based sensor selection method. The sensor selection method based on the greedy algorithm is described in detail in Section~\ref{sec:sensor}.} In addition, Eq.~\eqref{eq:observation_org} can be written as follows using the ROM of $\mathbf{x}$:

\begin{align}
    \mathbf{x}_p&\approx\mathbf{H}\mathbf{U}_r\mathbf{z+v} \nonumber \\
              &=\mathbf{C}\mathbf{z+v},
    \label{eq:observation}
\end{align}
where the columns of $\mathbf{U}_r\in\mathbb{R}^{n\times r}$ are the spatial modes of $\mathbf{x}$ and $\mathbf{z}\in\mathbb{R}^r$ are those mode coefficients. Here, $\mathbf{C}\in\mathbb{R}^{p\times r}$ is the measurement matrix which is the product of the sensor location matrix $\mathbf{H}$ and the sensor candidate matrix $\mathbf{U}$. The unknown variable in Eq.~\eqref{eq:observation} is only the mode coefficients $\mathbf{z}$. The estimated mode coefficients $\tilde{\mathbf{z}}$ can be obtained by the pseudo-inverse operation follows:
\begin{align}
    \tilde{\mathbf{z}}=\mathbf{C}^\dagger\mathbf{x}_p,
    \label{eq:LS_estimation_org}
\end{align}
where $\circ^\dagger$ indicates the pseudo-inverse operation. Hence, the estimated mode coefficients $\tilde{\mathbf{z}}$ can be calculated as follows by the least square estimation (LSE) :
\begin{align}
    \tilde{\mathbf{z}}=\left\{
    \begin{array}{ll}
    \mathbf{C}^{\mathsf T}\left(\mathbf{C}\mathbf{C}^{\mathsf T}\right)^{-1}\mathbf{x}_p, & p\leq r,\\
    \left(\mathbf{C}^{\mathsf T}\mathbf{C}\right)^{-1}\mathbf{C}^{\mathsf T}\mathbf{x}_p, & p>r.
    \label{eq:LS_estimation}
    \end{array}
    \right.
\end{align}
Then, the estimated full state can be obtained as the following simple multiplication:
\begin{align}
    \tilde{\mathbf{x}}=\mathbf{U}_r\tilde{\mathbf{z}}.
    \label{eq:reconst}
\end{align}

\subsection{Optimization of locations of observation sites}\label{sec:sensor}
Measurements are typically performed at specific points using discretely installed sensors. Therefore, it is necessary to maximize the information obtained with the sensors as much as possible. In some cases, it is better to use all available sensors, but in other cases, it is better to use fewer sensors for estimation from the perspective of the signal-to-noise ratio, processing time, and so on. \reviewerA{The sensor selection method is useful for also optimal design as one for laying out new networks or adding new stations to an already existing network. However, the objective of the present paper is to propose seismic wavefield reconstruction based on data-driven ROM, and therefore, site selection is only a tool in the present paper. In a real application, many observation sites are available, including far-field ones. The site selection in the context of the wavefield reconstruction is a selection of observation sites beforehand for each hypocentral region.}

The problem that optimizes sensor locations is called the sensor selection/placement problem and is formulated as a combinatorial optimization problem, which is known as a nondeterministic polynomial time (NP)-hard problem. In the present study, we employed the D-optimality-based greedy method for vector measurements (DG-vec) proposed by \citet{saito2021data} to select appropriate observation sites when the number of observation sites is reduced. This method is adapted to optimization of the sensor locations for vector measurement that the multiple component signals are measured by each sensor.  \reviewerB{Optimization of sensor locations was conducted based on the D-optimal design of experiment \citep{atkinson2007optimum}. The aim of DG-vec method is to find a suboptimal sensor set that minimizes the volume of the error ellipsoid in the linear inverse problem. Minimizing the estimation error is realized by maximizing the determinant of the error covariance matrix of the estimated latent variables by the least squares estimation and those true values.} The sensor candidate matrix was $\mathbf{U}_r$ in the present study. \reviewerB{This process corresponds to constructing measurement matrix $\mathbf{C}$ from the sensor candidate matrix $\mathbf{U}_r$ by constructing the sensor location matrix $\mathbf{H}$ in Eq.~\eqref{eq:observation} ($\mathbf{C}=\mathbf{HU}_r$). The sensor locations, which are the location of the unity in the row vectors of $\mathbf{H}$, are iteratively selected while adding the row of $\mathbf{H}$, and the following objective function is maximized in each single-sensor subproblem:}

\begin{align}
    f_{\mathrm{D}}\,=\,\left\{
    \begin{array}{cc}
        \mathrm{det}\,\left(\mathbf{CC}^{\mathsf T}\right), & p\le r, \\
        \mathrm{det}\,\left(\mathbf{C}^{\mathsf T}\mathbf{C}\right), & p>r.
    \end{array}\right.\label{eq:obj_det}
\end{align}

\noindent It should be noted that objective function for site selection in the present study was carried out by modified objective function of Eq.~\eqref{eq:obj_det} as follows for further robust computation \citep{shamaiah2010greedy,saito2021determinant}:
\begin{align}
    &f_{\mathrm{D}}=\mathrm{det}\,\left(\mathbf{C}^{\top}\mathbf{C}+\epsilon\mathbf{I}\right).
        \label{eq:obj_det2}
\end{align}

\noindent Here, $\epsilon$ and $\mathbf{I}$ are the sufficiently small positive real number and identity matrix, respectively. This objective function is available for both condition of $p\leq r$ and $p>r$ and robust, but it requires the hyperparameter $\epsilon$. The value of $\epsilon$ was set to $10^{-10}$ in the present study.

The matrices $\mathbf{U}_r$ and the vectors $\mathbf{z}$ in Eq.~\eqref{eq:observation} correspond to the sensor candidate matrix and the latent variables, respectively, in the sensor selection and full-state  recovery problems. The matrix $\mathbf{U}_r$ is usually a tall-and-skinny matrix (i.e., $n>>r$) in a practical problem because $n$ corresponds to the degrees of freedom in the spatial direction of the full state. The system Eq.~\eqref{eq:observation} represents the problem of choosing $p$ observations from $n$ state variables generated by the nondynamical system with $r$ latent variables. Although the sensor candidate corresponding to the available observation site in the present study is only $n=50$, the available observation sites will possibly become $\mathcal{O}\left(10^3\right)$--$\mathcal{O}\left(10^6\right)$ when it becomes possible to use vibrometer data implemented on buildings and utilities for wavefield reconstruction in the future. In such a case, the set of observation sites used for the wavefield reconstruction should be effectively reduced for fast and accurate reconstruction. The selection of observation sites used for wavefield reconstruction of earthquakes that occurred in each hypocentral region can be conducted beforehand using a sensor optimization technique.

\section{Numerical experiment}\label{sec:experiment}
\subsection{Problem settings}\label{sec:prob}
\begin{figure*}
\centering
\includegraphics[width=14.0cm]{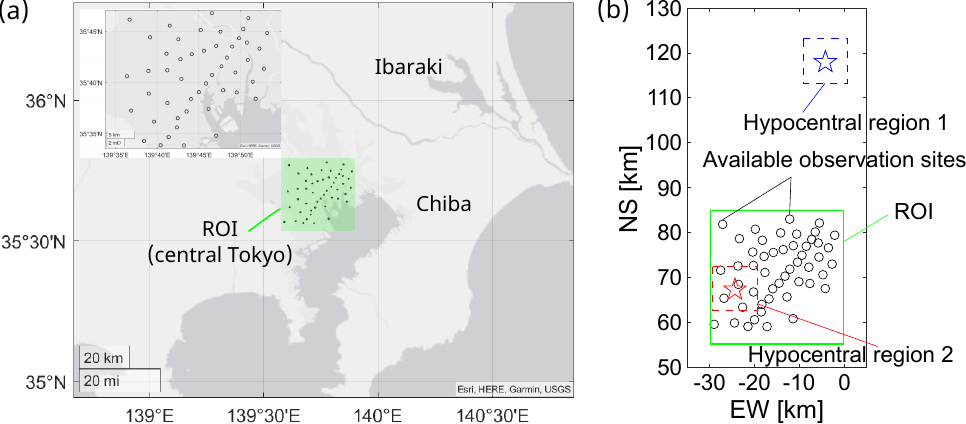}
\caption{(a) The Kanto area, showing the locations of a subset of the MeSO-net stations (circles). (b) Locations of the hypocentral regions and observation sites.}
\label{fig:problem_source}
\end{figure*}

The target area for seismic wavefield reconstruction in the present study is the metropolitan area of Tokyo, Japan. In this area, a dense seismological network named the Metropolitan Seismic Observation network (MeSO-net) has been in operation since 2007. The MeSO-net comprises 296 accelerometers, located at intervals of several kilometers, that continuously record seismograms at a sampling rate of 200~Hz \citep[e.g.,][]{hirata2009outline,sakai2009distribution}.
The present numerical experiment demonstrates the reconstruction of the spatially continuous wavefield of the region of interest (ROI) from a discretely measured waveform with the data-driven ROM, as shown in Figure~\ref{fig:problem_source}. Numerical simulations are carried out in advance for generating the synthetic dataset, and a low-dimensional model of the wavefield in the ROI is constructed. It is assumed that earthquakes occur in the assumed epicenter region. The ROI of the present experiment was the Tokyo 23 Ward area, which is located in the central Tokyo. We calculated the wavefield in the ROI ($55\leq\rm{NS~[km]}\leq85$, $-30\leq\rm{EW~[km]}\leq0$) that is discretised with an equally-spaced Cartesian grid, and the number of grid points was $n_x=N_{\rm NS}\times N_{\rm EW}=50\times50$. Here, $N_{\rm NS}$ and $N_{\rm EW}$ are the number of grid points in the NS and EW directions. The available observation sites exist in the ROI (green square in Figure~\ref{fig:problem_source}), and the number of points was $n_y=50$. These observation sites for the wavefield reconstruction were the subset of the MeSO-net observatories that actually exist in the Kanto region in Japan. Here, the location of the observation sites is not a subset of the grid points, i.e., the observation sites are not exactly on the grid points. Therefore, the formulation of the full-state recovery described in Section~\ref{sec:full-data_recovery} should be modified. This will be discussed in Section~\ref{sec:full-data_recovery_mod}.

\begin{figure}
\centering
\includegraphics[width=4.5cm]{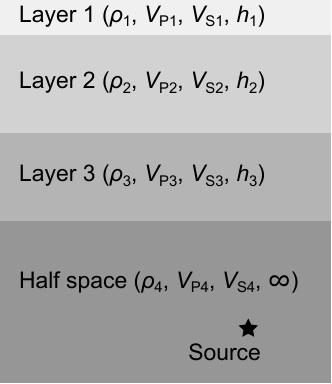}
\caption{Subsurface structure model.}
\label{fig:subsurface}
\end{figure}

Figure~\ref{fig:subsurface} depicts the assumed subsurface structure consisting of horizontally-stratified three layers, which models sedimentary basins. The model parameters of each layer are listed in Table~\ref{tab:ModelParam}. These layers are lying on a bedrock, which is modeled as a half-space. This subsurface model is adequate for the Kanto basin for the wavelengths of seismic waves considered in the present experiment. The earthquake source is assumed to locate in the half-space. Two hypocentral regions in which a number of earthquakes of similar type actually occur are assumed, i.e., the hypocentral region 1 in the southern part of Ibaraki Prefecture and the hypocentral region 2 in Tokyo 23 Wards. In the case of the hypocentral region 1, the center of the earthquake source region is the point where the magnitude 5.6 earthquake actually occurred on September 16, 2014. In the case of the hypocentral region 2, the center of the earthquake sources region is the point where the magnitude 4.1 earthquake actually occurred on May 14, 2021. Hypocenter solutions are shown in Table~\ref{tab:HypoSolution}. \reviewerB{The problem was simplified by assuming that earthquakes would occur somewhere in these regions surrounded by a blue or red dashed square and that the mechanism and direction of the fault slip would be constant for each hypocentral region.} The earthquake sources were uniformly distributed in $\left(\rm{NS},\rm{EW},\rm{UD}\right)=\left(118.0\pm5,\,-4.3\pm5,\,47.0\pm5\right)$~km (hypocentral region 1) or $\left(\rm{NS},\rm{EW},\rm{UD}\right)=\left(67.3\pm5,\,-24.4\pm5,\,72.0\pm5\right)$~km (hypocentral region 2). 

\begin{table}
\caption{Hypocenter solution of the standard case.}
\label{tab:HypoSolution}
\begin{tabular}{ll}
Hypocentral region 1 (Northern Kanto) & \\ \hline
Epicenter location (NS, EW, UD) [km] & 117.9655, -4.2204, 47.0  \\
Strike, rake, dip [deg] & 28, 118, 254 \\
Slip [m] & 0.4 \\ \hline
Hypocentral region 2 (Tokyo) & \\ \hline
Epicenter location (NS, EW, UD) [m] & 67.2743, -24.3698, 72.0 \\
Strike, rake, dip [deg] & 341, 160, 41 \\
Slip [m] & 0.06 \\ \hline
\end{tabular}
\end{table}

\begin{table}
\caption{Average value of the subsurface model parameters \citep{koketsu2008progress,koketsu2012japan}.}
\label{tab:ModelParam}
\begin{tabular}{lllll}
 & $\rho_i$ [g/cm$^3$] & $V_{{\rm P}i}$ [km/s]& $V_{{\rm S}i}$ [km/s]& $h_i$ [km] \\ \hline
Layer 1 & 1.95 & 1.8 & 0.5 & 0.4 \\
Layer 2 & 2.15 & 2.4 & 1.0 & 1.1 \\
Layer 3 & 2.3 & 3.2 & 1.7 & 1.0 \\
Half space (bedrock) & 2.7 & 5.8 & 3.4 & $\infty$ \\ \hline
\end{tabular}
\end{table}

\subsection{Dataset}\label{sec:dataset}
The waveform at each grid point for constructing the ROM and at each available observation site was obtained as the theoretical seismic waveform in the frequency domain. The theoretical waveforms were calculated using the code developed by \citet{hisada1995efficient}. This code calculates theoretical waveforms numerically based on the discrete wavenumber method \citep{bouchon1981simple}. The seismic waveforms are represented as wavenumber integrals of Green's functions. By introducing the reflexion and transmission matrix \citep{luco1983green,hisada1994efficient}, the Green's functions of layered half-space are calculated without numerical instabilities, even at higher frequencies. Therefore, theoretical waveforms are obtained with less computational cost, compared to the finite difference method or finite element method, assuming a 1-D subsurface structure model and given source information. \reviewerB{There are differences between a real wavefield and a synthesized one using a 1-D velocity structure depending on the underground structure and frequency band. This point is one of the major issues in applying the present framework to real datasets, but the influence of the difference in the velocity structure can be reduced by using the high-fidelity simulation with a 3-D velocity structure. In addition, the velocity structure itself can also be included as a parameter when constructing the ROM. However, high-fidelity simulation requires a large computational cost, and the objective of the present paper is to propose a seismic wavefield reconstruction framework based on compressed sensing using data-driven ROM. Thus, the numerical simulation with a 1-D subsurface structure model was employed in the present study. It should be noted that the waveform calculated using a 1-D velocity structure in the low-frequency band (approximately lower than 0.2~Hz) shows good agreement with that calculated using a 3-D velocity structure. Therefore, reconstruction of low-frequency waves, such as direct waves from epicenters with a period longer than 10 seconds that can damage large structures in a metropolitan area, is possible even if the simulation uses a 1-D subsurface structure.}

The dataset was generated by many calculations with different conditions. The length of the calculated waveform was 40.96~s, and the sampling frequency was 100~Hz. Calculations were conducted with $m=1008$ parameter sets. The location of the seismic source, the propagation velocities of the P and S waves in each layer ($V_{{\rm P}i}$ and $V_{{\rm S}i}$, where $i=1,2,3$), and the thickness of each layer $h_i$ were changed randomly, and other parameters were fixed. \reviewerA{Therefore, the source mechanism was fixed in the generation of training data to simplify the problem, and thus, the constructed ROM cannot reproduce the influence of the source mechanism that dramatically changes the character of the wavefield. This problem can be avoided by including the source mechanism as an additional parameter when generating the training data. In that case, the required training data and rank of the model are considered to increase.} The variation in the locations of the seismic sources was according to the uniform distribution in the range of $\pm5$~km in each direction from the center location of the hypocentral region, as described in Section~\ref{sec:prob}. The stochastic variables $V_\mathrm{Pi}$, $V_\mathrm{Si}$, $h_i$ varied in accordance with a multivariate normal distribution that has a diagonal covariance matrix. The average of each parameter was shown in Table~\ref{tab:ModelParam}, and the standard deviation was 10\% of values of each parameter. Here, the average value of the subsurface model parameter is based on the subsurface structure model of the Japan Integrated Velocity Structure Model by \cite{koketsu2008progress,koketsu2012japan}.

The waveforms observed at each point are obtained as Fourier spectra. It has 205 frequency components for each three spatial directions (NS, EW, UD). Hence, the measurement data at each point have $s=1230$ components considering each frequency component consisting of the real and imaginary parts. The waveform data for each parameter set at the grid points and the available observation sites were shaped into column vectors $\mathbf{x}\in\mathbb{R}^{(s\times n_x)\times1}$ and $\mathbf{y}\in\mathbb{R}^{(s\times n_y)\times1}$, respectively. The structure of the data vectors is $\mathbf{x}=\left[\mathbf{\xi}_1^1\,\,\mathbf{\xi}_2^1\,\cdots\,\mathbf{\xi}_n^1\,\cdots\,\mathbf{\xi}_1^s\,\,\mathbf{\xi}_2^s\,\cdots\,\mathbf{\xi}_n^s\right]^{\mathsf T}$ and $\mathbf{y}=\left[\mathbf{\eta}_1^1\,\,\mathbf{\eta}_2^1\,\,\cdots\,\mathbf{\eta}_n^1\,\cdots\,\mathbf{\eta}_1^s\,\,\mathbf{\eta}_2^s\,\,\cdots\,\mathbf{\eta}_n^s\right]^{\mathsf T}$, where $\xi$ and $\eta$ are the real or imaginary parts of a certain frequency of Fourier spectra obtained at a certain point. The data matrices for the grid points $\mathbf{X}=\left[\mathbf{x}_1\,\,\mathbf{x}_2\,\cdots\,\mathbf{x}_m\right]$ and the available observation sites $\mathbf{Y}=\left[\mathbf{y}_1\,\,\mathbf{y}_2\,\cdots\,\mathbf{y}_m\right]$ were constructed.

\begin{figure}
\centering
\includegraphics[width=6.5cm]{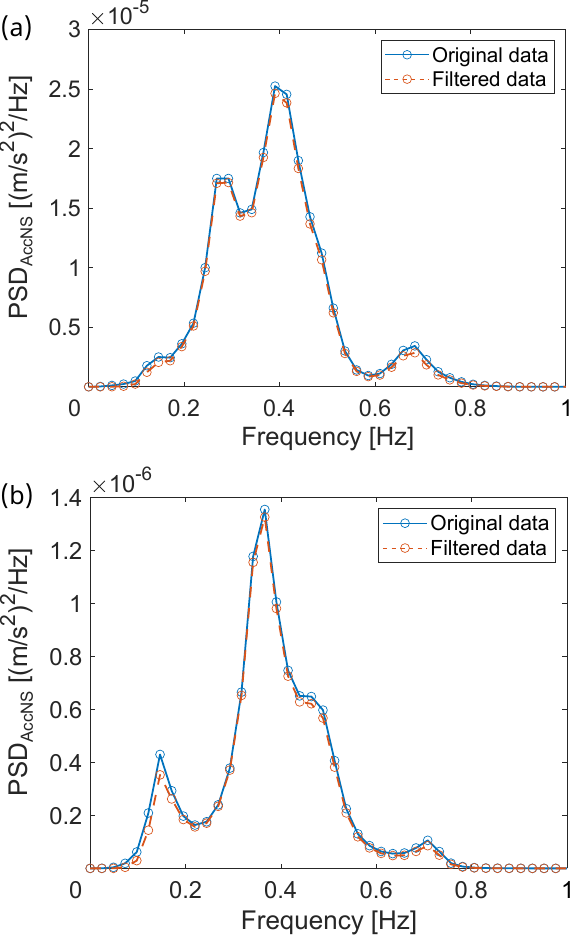}
\caption{Comparison of frequency spectra of the original data and the filtered data at the center of the ROI: (a) hypocentral region 1; (b) hypocentral region 2.}
\label{fig:PSDorg_25-50}
\end{figure}

The present simulation was conducted in the frequency range less 5~Hz, and the second-order Butterworth filter of the frequency band $0.1\leq f \leq 1.0$ was applied to the obtained datasets $\mathbf{X}$ and $\mathbf{Y}$. The comparison of the power spectral density (PSD) of the original data and filtered data at the center of the RIO is shown in Figure~\ref{fig:PSDorg_25-50}. The filtered dataset was further used for reduced-order modeling, selection of observation sites, and the wavefield reconstruction experiments.

\subsection{Reduced-order modelling and full-state recovery}\label{sec:full-data_recovery_mod}
The data-driven ROM of the wavefield was obtained by the truncated SVD of the data matrix $\mathbf{X}$ \reviewerB{which contains the Fourier spectra of the wavefield obtained on the grid},

\begin{align}
\mathbf{X}\approx\mathbf{U_x}\mathbf{S_x}\mathbf{V_x}^{\mathsf T}=\mathbf{U_x}\mathbf{Z_x}.
\end{align}

\noindent Here, obtained matrices $\mathbf{U_x}$, $\mathbf{S_x}$, and $\mathbf{V_x}$ consist of leading-$r$ modes, as described in Eq.~\eqref{eq:ROM}. In the present study, the number of retained modes $r$ for the ROM was determined based on the model error $\epsilon_{\rm model}$, and the number of retained modes $r$ was set so that the model error would be $\epsilon_{\rm model}\approx10$\%. \reviewerA{It should be noted that the model error becomes large when the number of retained modes $r$ increases.} The definition of the model error is as follows:

\begin{align}
\epsilon_{\rm model} = \frac{\|\mathbf{X}-\mathbf{U_x Z_x}\|_{\rm F}}{\|\mathbf{X}\|_\mathrm{F}},
\label{eq:modelerr}
\end{align}

\noindent where the notation $\|\circ\|_\mathrm{F}$ is the Frobenius norm that is the matrix norm as the square root of the sum of the absolute squares of the matrix elements. This error is due to mode truncation and is a lower bound of the reconstruction error of the present framework. 
The subscript $\circ_\mathbf{x}$ indicates the variable for the grid. Typically, the positions of the observation site do not match the grid points. Therefore, we determine the relationship between the measurement data $\mathbf{Y}$ obtained by the available observation sites and the mode coefficient of the ROM of the wavefield $\mathbf{Z_x}$ by linear regression as follows:

\begin{align}
    \mathbf{Y}&\approx\mathbf{U_y}\mathbf{Z_x}, \nonumber \\
    \mathbf{U_y}&=\mathbf{Y}\mathbf{V_x}\mathbf{S_x^{-1}},
\end{align}

\noindent where the subscripts $\circ_\mathbf{y}$ indicate the variables for the observation sites.

When estimating the wavefield from the measurement data obtained at the observation sites $\mathbf{y}$, the measurement equation described in Eq.~\eqref{eq:observation} becomes as follows: 

\begin{align}
\mathbf{y}_p&=\mathbf{H}\mathbf{y}+\mathbf{w}, \nonumber \\
            &\approx\mathbf{H}\mathbf{U_yz_x}+\mathbf{w}, \nonumber \\
            &=\mathbf{C_y}\mathbf{z_x}+\mathbf{w},
\label{eq:observation2}
\end{align}

\noindent where $\mathbf{y}_p\in\mathbb{R}^{(s\times p)\times1}$ is the observation vector which consists of measured data obtained by $p$-activated observation sites selected among $n_y$ available sites, and $\mathbf{w}\in\mathbb{R}^{(s\times p)\times 1}$ is observation noise.
Then, the estimated mode coefficients $\tilde{\mathbf{z}}$ are obtained by the pseudo-inverse operation

\begin{align}
\tilde{\mathbf{z}}_{\mathbf{x}}&=\mathbf{C_y}^{\dagger}\mathbf{y}_p,
\end{align}

\noindent and the estimated wavefield on the grid $\tilde{\mathbf{x}}$ can be obtained as follows:

\begin{align}
\tilde{\mathbf{x}}&=\mathbf{U_x}\tilde{\mathbf{z}}_{\mathbf{x}}.
\end{align}

\section{Results and Discussion}\label{sec:results}
In this section, the results of seismic wavefield reconstruction using the present framework are provided. The results of the selection of observation points using the DG-vec method are firstly described. The characteristics of the present framework are discussed based on the reconstruction results using noise-free and noise-contaminated observations in Section~\ref{sec:wonoise} and \ref{sec:wnoise}, respectively. In the numerical experiment, the generated dataset of the wavefield was split into five segments, and model construction was carried out using 80\% of the generated dataset, and the other 20\% of the generated dataset was used as test data. The five-fold cross-validation was performed for the evaluation of reconstruction error, and average error and standard deviation were calculated. 

\subsection{Priority of observation site}
\begin{figure}
\centering
\includegraphics[width=6.5cm]{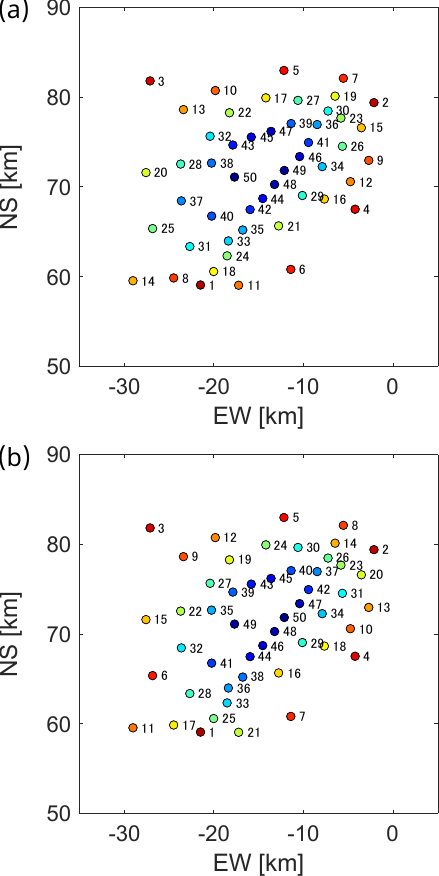}
\caption{Priority of observation sites in the selection of observation sites using the DG-vec method: (a) hypocentral region 1; (b) hypocentral region 2.}
\label{fig:observation_sites}
\end{figure}

The priority of the selected observation sites by DG-vec method for the hypocentral regions 1 and 2 is displayed in Figure~\ref{fig:observation_sites}. \reviewerB{In the case of the hypocentral regions 1 and 2, the numbers of retained modes for ROM of the wavefield $r$ were $r=85$ and $r=57$, respectively.} In the present study, the DG-vec method, which is based on the greedy algorithm, was adopted for the selection of observation sites. \reviewerB{In the present study, the reconstruction of the wavefield are conducted in the frequency domain based on the Fourier spectra of the observed waveform at the observation sites. Hence, the site selection was carried out in the frequency domain. The sensor candidate matrix is $\mathbf{U_y}$ in Eq.~\eqref{eq:observation2}).}
This algorithm iteratively selects observation sites in sequence to maximize the objective function in each single-sensor subproblem. Therefore, the order of selected observation sites by the greedy algorithm can be regarded as the priority of the observation sites. Each observation site in Figure~\ref{fig:observation_sites} is labeled with the priority of the observation sites (order of selection in the greedy algorithm). The color of each observation site also corresponds to the priority of observation sites. In both cases, the observation sites located near the outer edge of the set of the observation sites are preferentially selected, and the observation sites are selected so as to the  ROI is contained inside the region of the selected observation sites as much as possible. Although the order of the selected observation sites is slightly different, the trend is consistent even if the hypocentral region was different, and there is no strong influence of the positional relationship between the hypocentral region and the ROI on the priority of the observation sites. This is because all observation sites are distributed inside the ROI, and \reviewerB{the signal was also sufficiently high in the present case. This trend is considered to be altered depending on the relation between the size of the ROI and the set of the observation sites. The difference in the signal intensity seems to affect the site selection, if the ROI is wide and the influence of decay of seismic waves is not negligible.}

\subsection{Reconstruction with noise-free observation}\label{sec:wonoise}
\begin{figure*}
\centering
\includegraphics[width=16.5cm]{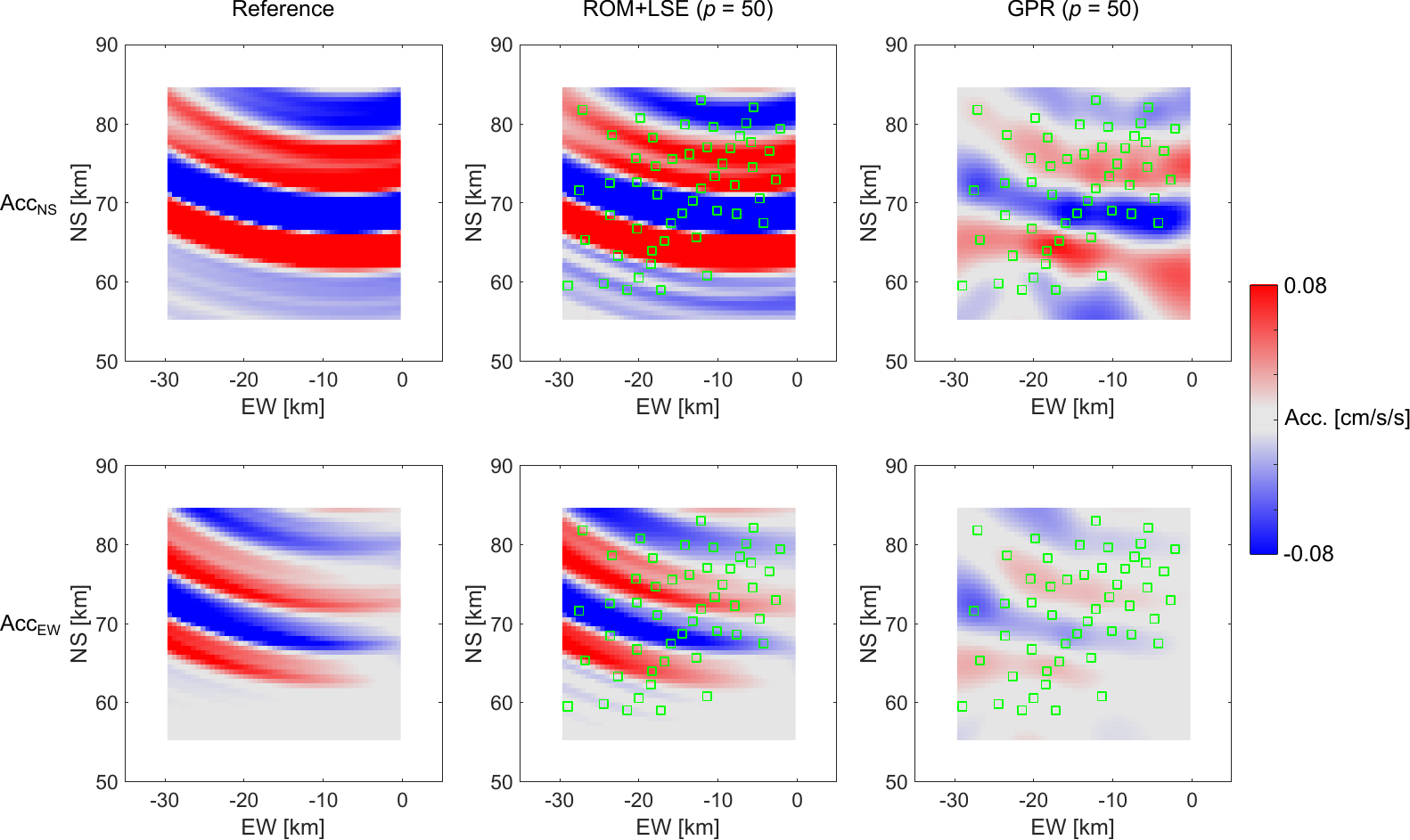}
\caption{Reconstructed wavefields obtained by the present method and the GPR-based method at $t=24.24$~s in the case of the hypocentral region 1 (w/o observation noise). [This figure appears in color in the online issue.]}
\label{fig:wavefieldGP_out_noisefree}
\end{figure*}

\begin{figure*}
\centering
\includegraphics[width=16.5cm]{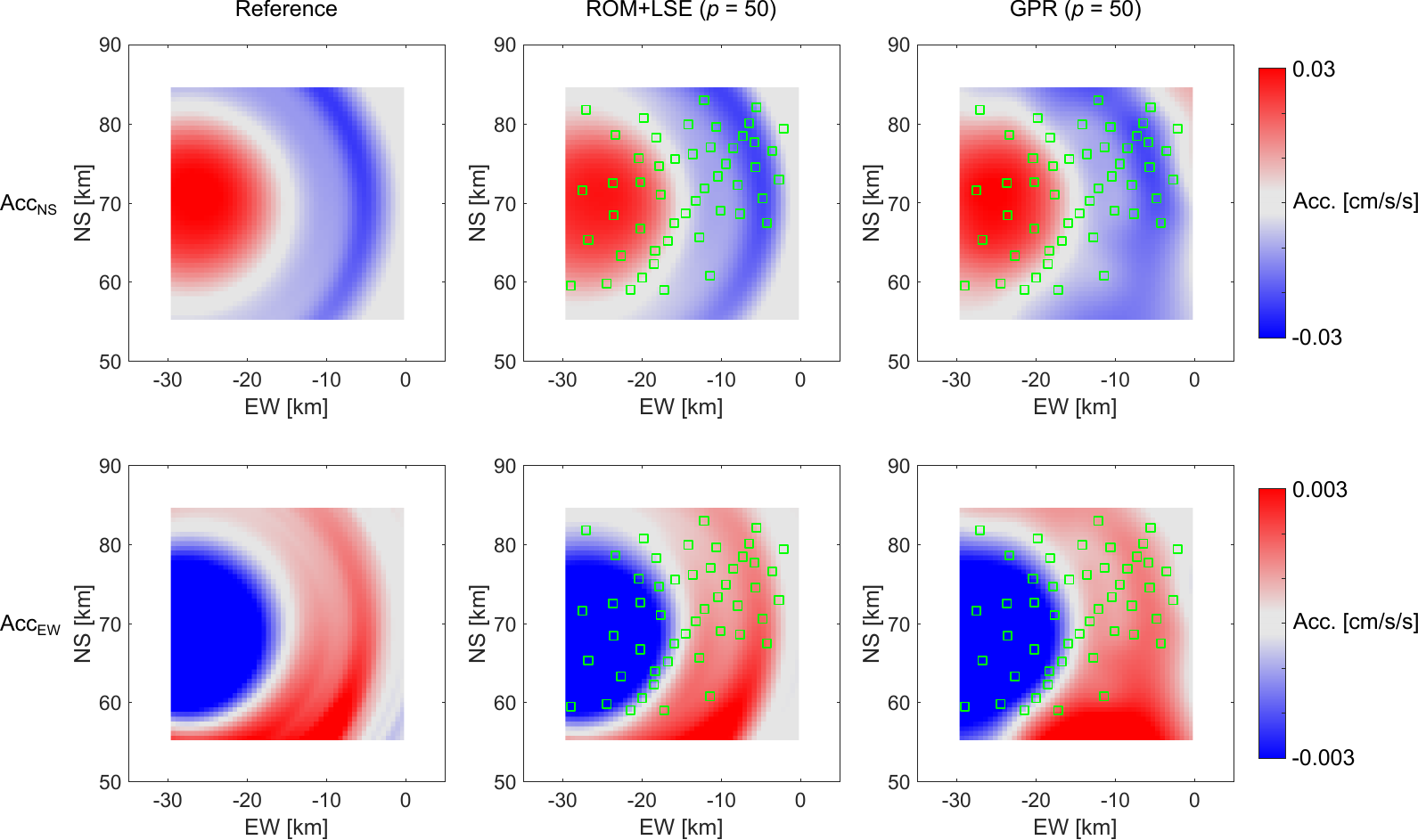}
\caption{Reconstructed wavefields obtained by the present method and the GPR-based method at $t=24.88$~s in the case of the hypocentral region 2 (w/o observation noise). [This figure appears in color in the online issue.]}
\label{fig:wavefieldGP_in_noisefree}
\end{figure*}

Figures~\ref{fig:wavefieldGP_out_noisefree} and \ref{fig:wavefieldGP_in_noisefree} show the snapshot of the reference data (filtered original data) and the reconstructed wavefield obtained by the present method (ROM+LSE) and the Gaussian process regression (GPR) at a certain time. \reviewerB{The cases for comparisons of the wavefield in Figures ~\ref{fig:wavefieldGP_out_noisefree} and \ref{fig:wavefieldGP_in_noisefree} were randomly selected. Therefore, the chosen cases are not necessarily the cases with the lowest reconstruction error for each method. It should be noted that not only the best performance, but also the variance in the estimation are important for wavefield reconstruction. The detailed comparisons are conducted based on the reconstruction error.}

\reviewerA{The green square symbols indicate the observation sites used for estimation, and the black circles are not used ones.} Distributions of the acceleration were estimated from the observed signal measured with all available observation sites. Although only the wavefields in horizontal direction are shown in Figures~\ref{fig:wavefieldGP_out_noisefree} and \ref{fig:wavefieldGP_in_noisefree}, the trend in the reconstructed wavedields in UD direction were similar to those in the horizontal directions. The observed signals are assumed to not be contaminated by noise. For reconstruction using the GPR-based method, the distribution of the Fourier spectra on the grid was estimated (spatially interpolated) from the Fourier spectra obtained at the observation sites. After reconstruction of the Fourier spectra, time series wavefields are calculated, the same as the ROM-based method. Model construction and estimation were conducted separately for the real and imaginary parts of each frequency in each direction. The squared exponential kernel was used as the kernel function. The noise variance and the parameters included in the kernel function, which are the parameters for the GPR-based method, were optimized for each component (frequencies and spatial directions) of the data by Bayesian optimization for only the first case of the dataset, and the optimized parameters were used for the other cases. \reviewerB{This Bayesian optimization of the hyperparameter was only conducted for the GPR-based method.} \reviewerA{It should be noted that the GPR-based method uses almost no information regarding the seismic wavefield, such as the relationship between frequency and wavelength and the basis of the wavefield. On the other hand, the ROM-based method uses effective prior information regarding the wavefield, such as the mode of the wavefield.}

The reconstructed wavefield obtained by the present method is almost the same as the reference data for both cases of the hypocentral regions 1 and 2. On the other hand, the quality of the reconstructed wavefield obtained by GPR is worse than that of the present method. In the case of the hypocentral region 1, the reconstructed wavefield captures the qualitative characteristics of the wavefields, but the reconstructed wavefield is strongly distorted. In the case of the hypocentral region 2, the quality of the reconstructed wavefield is relatively high, but distortion in the reconstructed wavefield is still strong in the region at around $({\rm EW},~{\rm NS}) = (-10,~55)$~km. This is because there are much fewer observation sites in this region compared to other regions, and thus, wavefield is estimated as extrapolation. On the other hand, the present method utilizes the spatial modes of the wavefield, and only an estimation of those coefficients based on the observed waveforms is needed for wavefield reconstruction. Hence, the wavefield can be estimated accurately, even though the region where there are fewer observation sites.

\begin{figure}
\centering
\includegraphics[width=6.5cm]{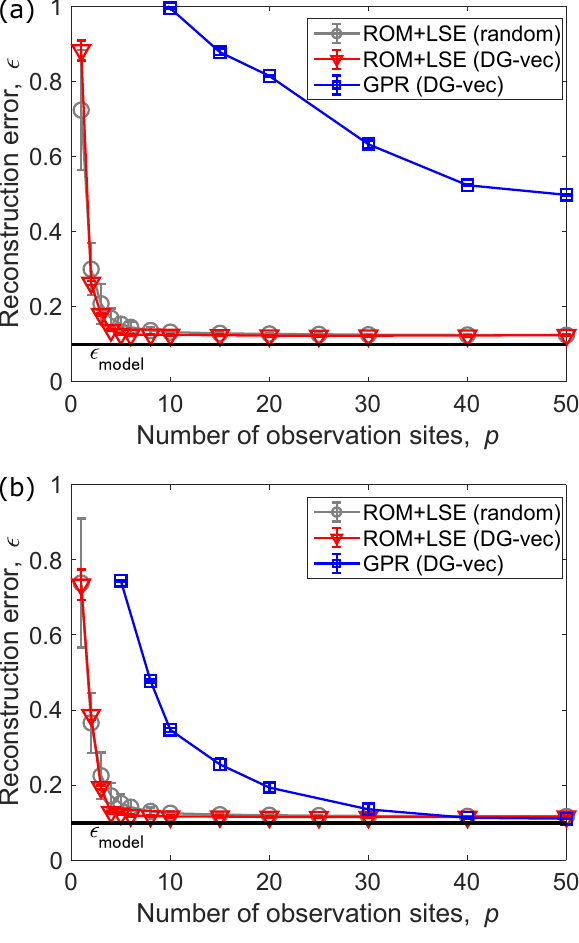}
\caption{Reconstruction error with respect to the number of observation sites (w/o observation noise): (a) hypocentral region 1; (b) hypocentral region 2.}
\label{fig:reconst_noisefree}
\end{figure}

The relation between the number of selected observation sites and the reconstruction error in the case of noise-free conditions is described in Figure~\ref{fig:reconst_noisefree}. Reconstruction was also conducted using the GPR-based method. The reconstruction error was calculated as follows:
\begin{align}
\epsilon = \frac{\|\mathbf{X}_{\rm test}-\mathbf{U_x\hat{Z}_x}\|_{\rm F}}{\|\mathbf{X}_{\rm test}\|_\mathrm{F}}.
\label{eq:recosterr}
\end{align}
\reviewerA{The error bars are displayed in Figure~\ref{fig:reconst_noisefree} based on the standard deviation of the reconstruction error obtained by five-fold cross-validation. However, those are smaller than the size of the symbol, except for the reconstruction using the proposed method with randomly selected observation sites.} In the case of random selection, 100 sets of the randomly generated observation site are generated, and the wavefield reconstruction was conducted. \reviewerB{The observation sites for reconstruction using fewer observation sites were selected by the DG-vec method for both proposed method and GPR-based method. This is because there is no published sensor optimization method for estimation using the GPR-based method. Although the observation sites selected by the DG-vec method are not optimized for the GPR-based method, the performance of the selected observation sites should be better than that of randomly selected observation sites.} 

The reconstruction error obtained by the GPR-based method is larger than that of the ROM-based method. Although the location of the observation sites is not optimized for reconstruction using the GPR-based method, the reconstruction error obtained by the ROM-based method is much smaller than that obtained by the GPR-based method, even though the randomly selected observation sites are used. 
\reviewerA{This is because that the ROM-based method uses the strong prior information of the wavefield, such as the mode of the wavefield. On the other hand, the GPR-based method uses almost no information regarding the seismic wavefield, such as the relationship between frequency and wavelength and the basis of the wavefield.}
The reconstruction error decreases as the number of selected observation sites increases and is asymptotic to model error $\epsilon_{\rm model}=0.1$. The reconstruction error rapidly decreases as the number of selected observation sites $p$ increases until $p=7$ for the ROM-based method. The reconstruction error can be reduced to the level of model error in the case of noise-free observation, even with the small number of selected observation sites. Furthermore, the difference between the reconstruction error obtained using the observation sites selected by random selection and the DG-vec method is small. This is because the problem is simple and the ROM-based method quite works well, and thus, sufficient information necessary for reconstruction can be obtained, even if the set of observation sites is not carefully selected. The effectiveness of the optimization of the location of observation sites can be seen in the case of noise-contaminated observations (Section~\ref{sec:wnoise}).

\begin{figure*}
\centering
\includegraphics[width=16.5cm]{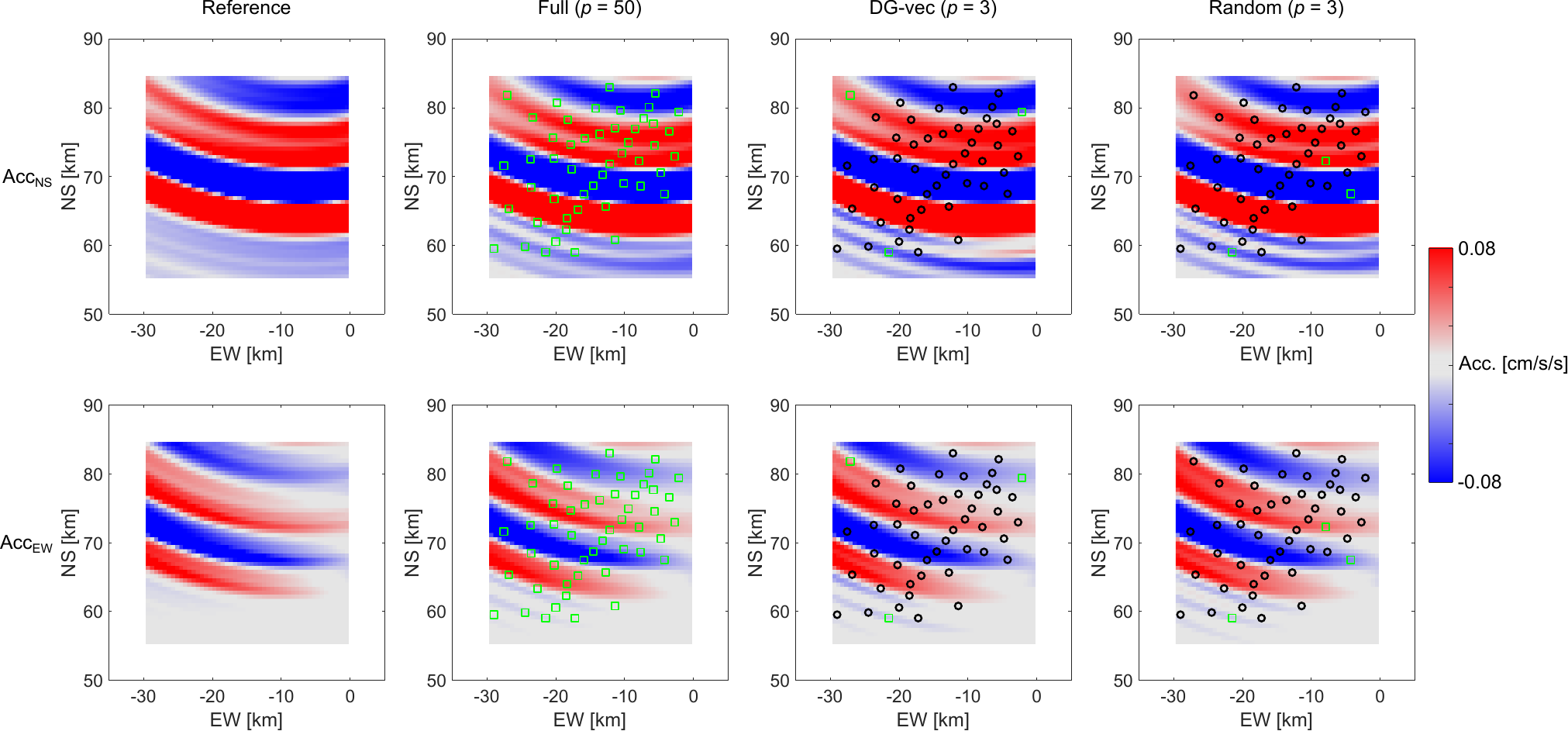}
\caption{Effect of the number of observation sites and selection method on the reconstructed wavefields at $t=24.24$~s in the case of the hypocentral region 1 (w/o observation noise). [This figure appears in colour in the online issue.]}
\label{fig:wavefield_out_noisefree}
\end{figure*}

\begin{figure*}
\centering
\includegraphics[width=16.5cm]{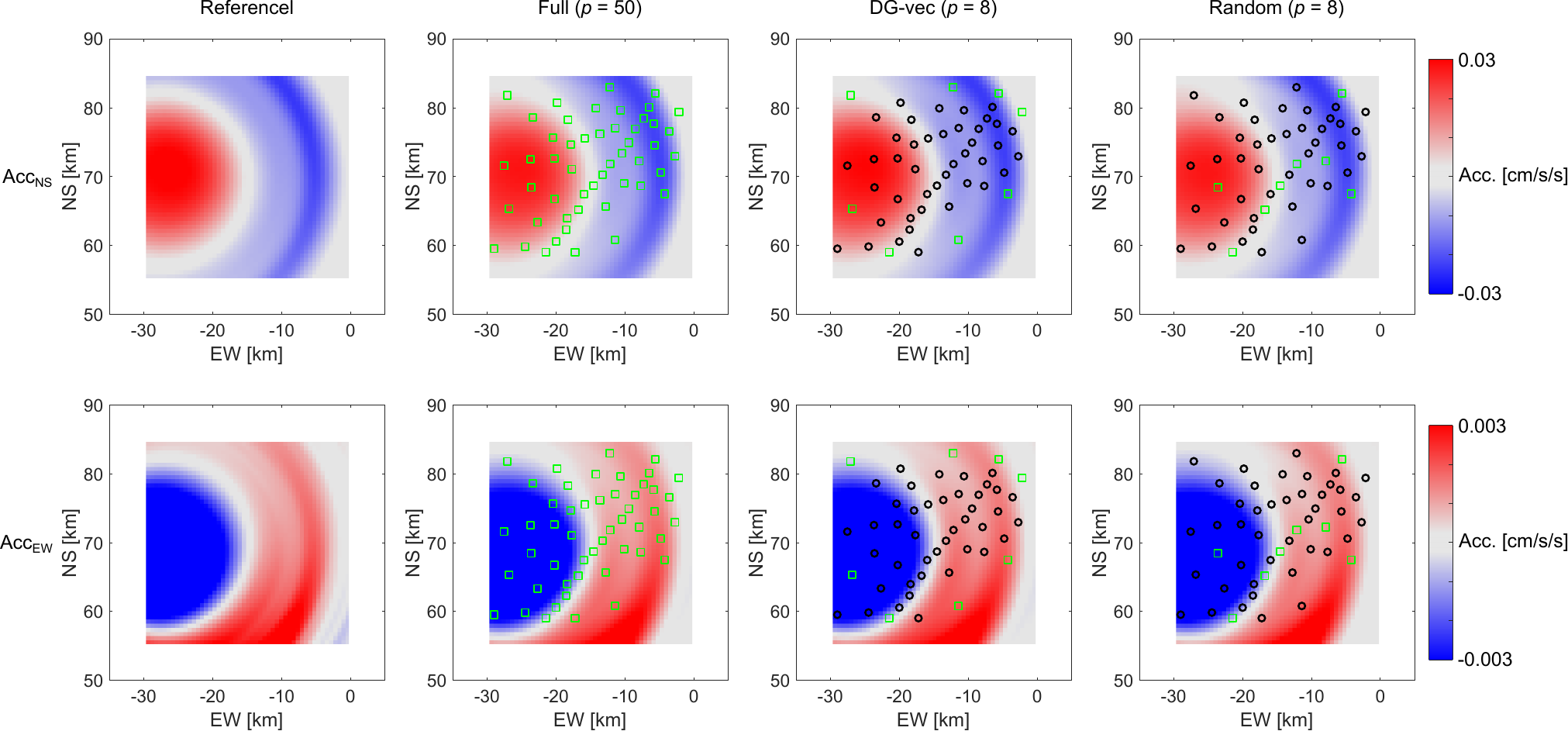}
\caption{Effect of the number of observation sites and selection method on the reconstructed wavefields at $t=24.88$~s in the case of the hypocentral region 2 (w/o observation noise). [This figure appears in colour in the online issue.]}
\label{fig:wavefield_in_noisefree}
\end{figure*}

Figures~\ref{fig:wavefield_out_noisefree} and \ref{fig:wavefield_in_noisefree} show the comparison of the reference and reconstructed wavefields at a certain time with a smaller number of observation sites. The distributions of the acceleration in the horizontal directions were estimated from the signal observed by the selected observation sites. The observation sites were selected randomly or by the DG-vec method. The number of selected observation sites was $p=3$ for the case of the hypocentral region 1 and was $p=8$ for the case of the hypocentral region 2. \reviewerA{These numbers of observation sites are the condition where the difference in the wavefield reconstructed with the observation sites selected by the random selection and the DG-vec method is large. The same conditions are also applied in the case of the reconstruction with noise-contaminated observations.} The reconstructed wavefields estimated from all observation sites are also displayed.

\noindent The reconstructed wavefield obtained by our framework is very similar to the reference data when all observation sites are used. The overall characteristics of the wavefield can be reproduced even when the number of observation sites is reduced. Therefore, reconstruction of the wavefield based on the ROM and discrete observation works well. 

In real applications, some observation sites may malfunction when an earthquake occurs, and the observed data at malfunctioning observation sites can give a negative influence on the wavefield reconstruction. Conducting wavefield reconstruction with multiple sets of observation sites and comparing the results is the simplest solution. For this validation, the set of observation sites for wavefield reconstruction is iteratively selected among observation site candidates, of which several observation sites are randomly masked. The difference in the performance of the set of observation sites can be minimized by using the optimization method of the observation sites, such as the DG-vec method, and the verification of the reconstructed wavefield can be conducted more properly than that using the sets of randomly selected observation sites. Also, the processing time for wavefield reconstruction can be reduced while reducing the degradation of the accuracy of the estimated wavefield by combining the optimization method of the observation sites.

\begin{figure}
\centering
\includegraphics[width=12.5cm]{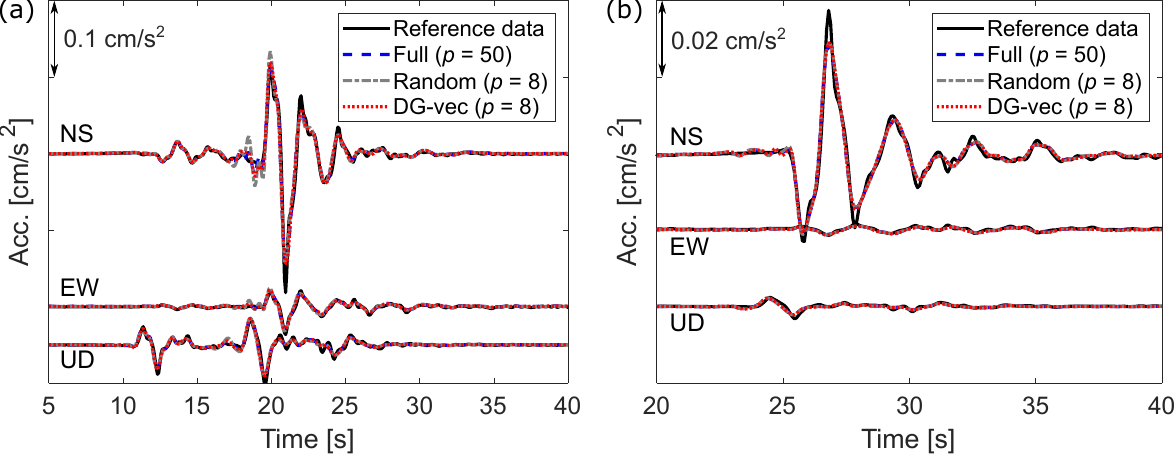}
\caption{\reviewerB{Reconstructed waveform in the center of the ROI (w/o observation noise): (a) hypocentral region 1 and (b) hypocentral region 2.}}
\label{fig:waveform_noisefree}
\end{figure}

\begin{figure}
\centering
\includegraphics[width=6.5cm]{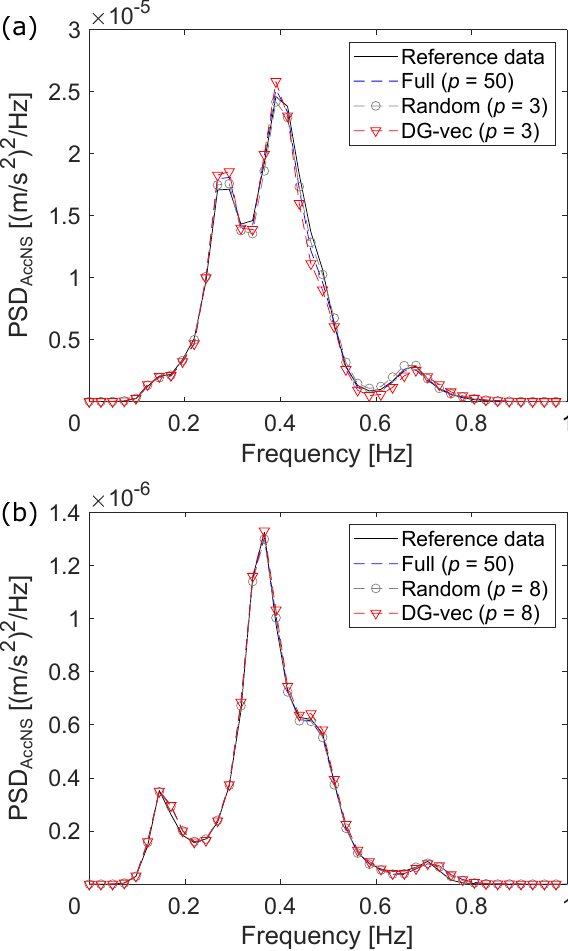}
\caption{Frequency spectra of the reconstructed data at the center of ROI (w/o observation noise): (a) hypocentral region 1; (b) hypocentral region 2.}
\label{fig:PSDreconst_25-50_noisefree}
\end{figure}

The accuracy of the reconstructed wavefield can be confirmed from the waveform and its frequency spectra, as shown in Figures~\ref{fig:waveform_noisefree} and \ref{fig:PSDreconst_25-50_noisefree}. Although 100 sets of the randomly generated observation site are generated, the frequency spectrum labeled as ''Random'' was that obtained by a certain set of observation sites. The waveforms are sampled at the center of the ROI, as the representative location. These figures illustrate that the reconstructed waveforms and those frequency spectra show good agreement with the reference data. However, the reconstructed waveforms are slightly smoothed and tend to have smaller amplitudes at the S-wave. These characteristics of the reconstructed waveform are due to the characteristics of the low-dimensional model. For the effectiveness of the selection of the observation sites, the reconstructed waveform with the eight observation sites selected by the DG-vec method is almost the same as that with 50 observation sites, which are all of the available observation sites. Although the difference is slight, on the other hand, the waveform reconstructed with the randomly selected eight observation sites for the hypocentral region 2 differs from that obtained with the 50 observation sites (e.g., at immediately before S-wave arrival).

\subsection{Reconstruction with noise-contaminated observation}\label{sec:wnoise}
\begin{figure}
\centering
\includegraphics[width=12.5cm]{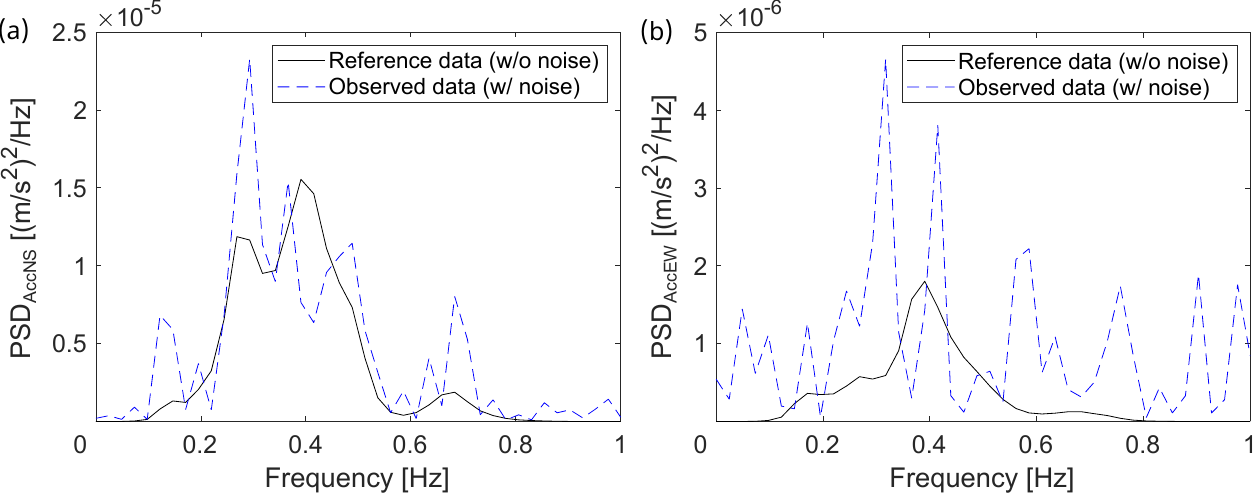}
\caption{\reviewerB{Frequency spectra of noise-free and noise-contaminated observation in the case of the hypocentral region 1.}}
\label{fig:PSDobserved_out_25-50}
\end{figure}

\begin{figure}
\centering
\includegraphics[width=12.5cm]{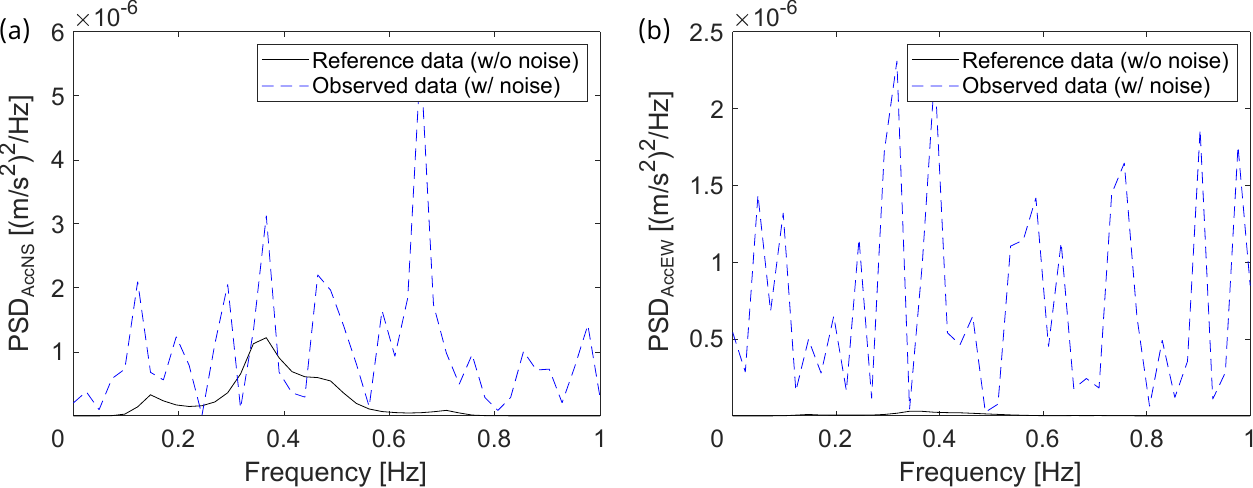}
\caption{\reviewerB{Frequency spectra of noise-free and noise-contaminated observation in the case of the hypocentral region 2.}}
\label{fig:PSDobserved_in_25-50}
\end{figure}

The results of the seismic wavefield reconstruction are provided in this section. Observation noise $\mathbf{w}$ was added to the observation data $\mathbf{y}_p$ for reconstruction. The amplitude of additive noise for each component of observation data (each entry of $\mathbf{y}_p$) follows a normal distribution $\mathbf{w}\sim\mathcal{N}(0,\sigma^2)$. \reviewerA{In the present study, the variance of the noise component was set to $\sigma^2=1.0\times10^{-8}$. This is consistent with adding the white noise, the mean of which is $\sqrt{2}\times10^{-4}$ [m/s/s], in the 
Fourier spectrum. Although the presented results are limited to the reconstruction error (see Figure~\ref{eq:reconst}), the reconstruction with other noise levels of $\sigma^2=2.5\times10^{-9}$ and $\sigma^2=2.5\times10^{-8}$ were conducted.} Figures~\ref{fig:PSDobserved_out_25-50} and \ref{fig:PSDobserved_in_25-50} show the frequency spectra of the noise-free and noise-contaminated observed signals. Note that the observation noise was added also to the observed signal in the UD direction. The noise level was set to high to emphasize the influence of the observation noise. Particularly, the noise level was severe in the case of the hypocentral region 2, and thus, the PSD of acceleration of the reference data is too small to visualize in Figure~\ref{fig:PSDobserved_in_25-50}.

\begin{figure*}
\centering
\includegraphics[width=16.5cm]{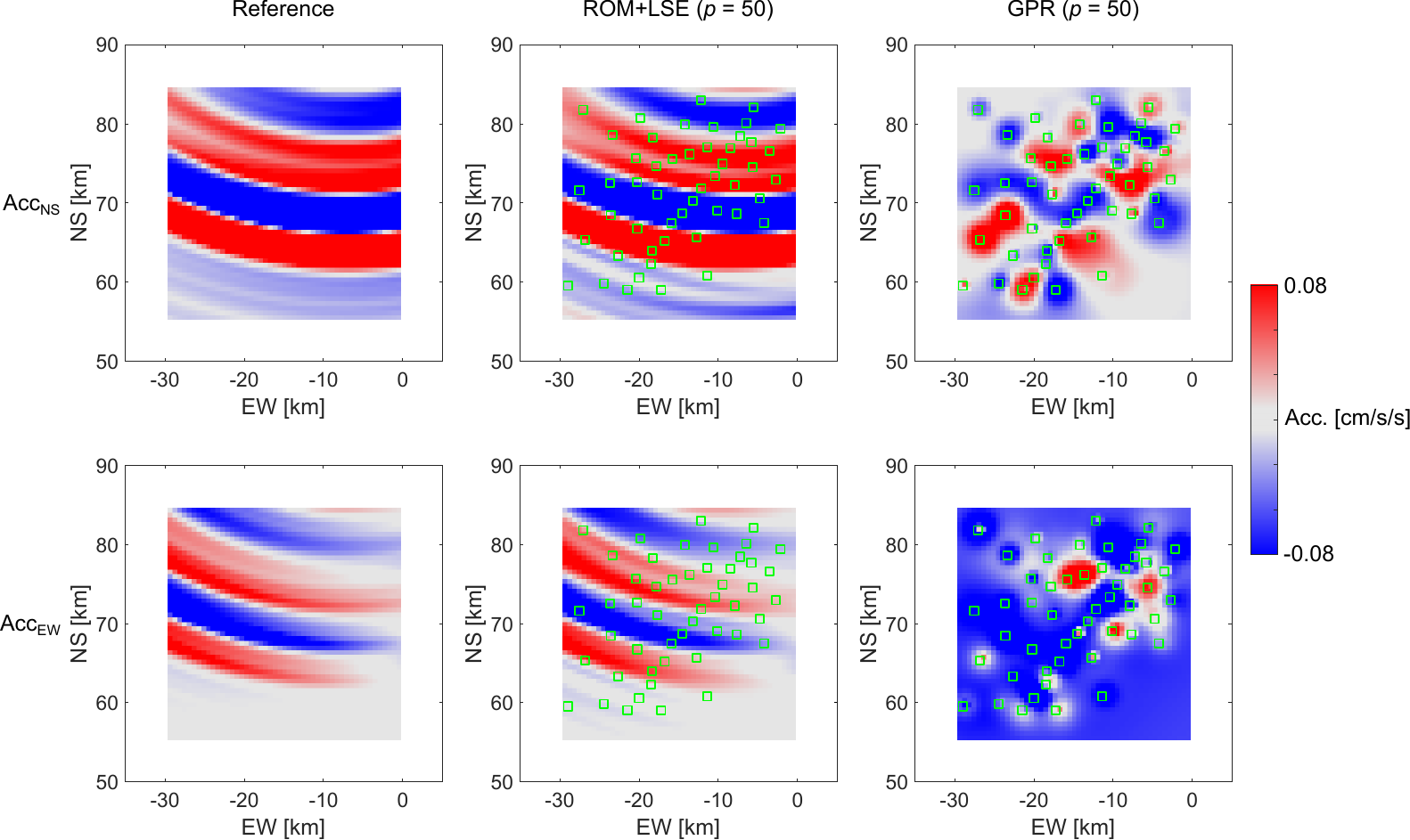}
\caption{Reconstructed wavefields obtained by the ROM-based method and the GPR-based method at $t=24.24$~s in the case of the hypocentral region 1 (w/ observation noise). [This figure appears in colour in the online issue.]}
\label{fig:wavefieldGP_out}
\end{figure*}

\begin{figure*}
\centering
\includegraphics[width=16.5cm]{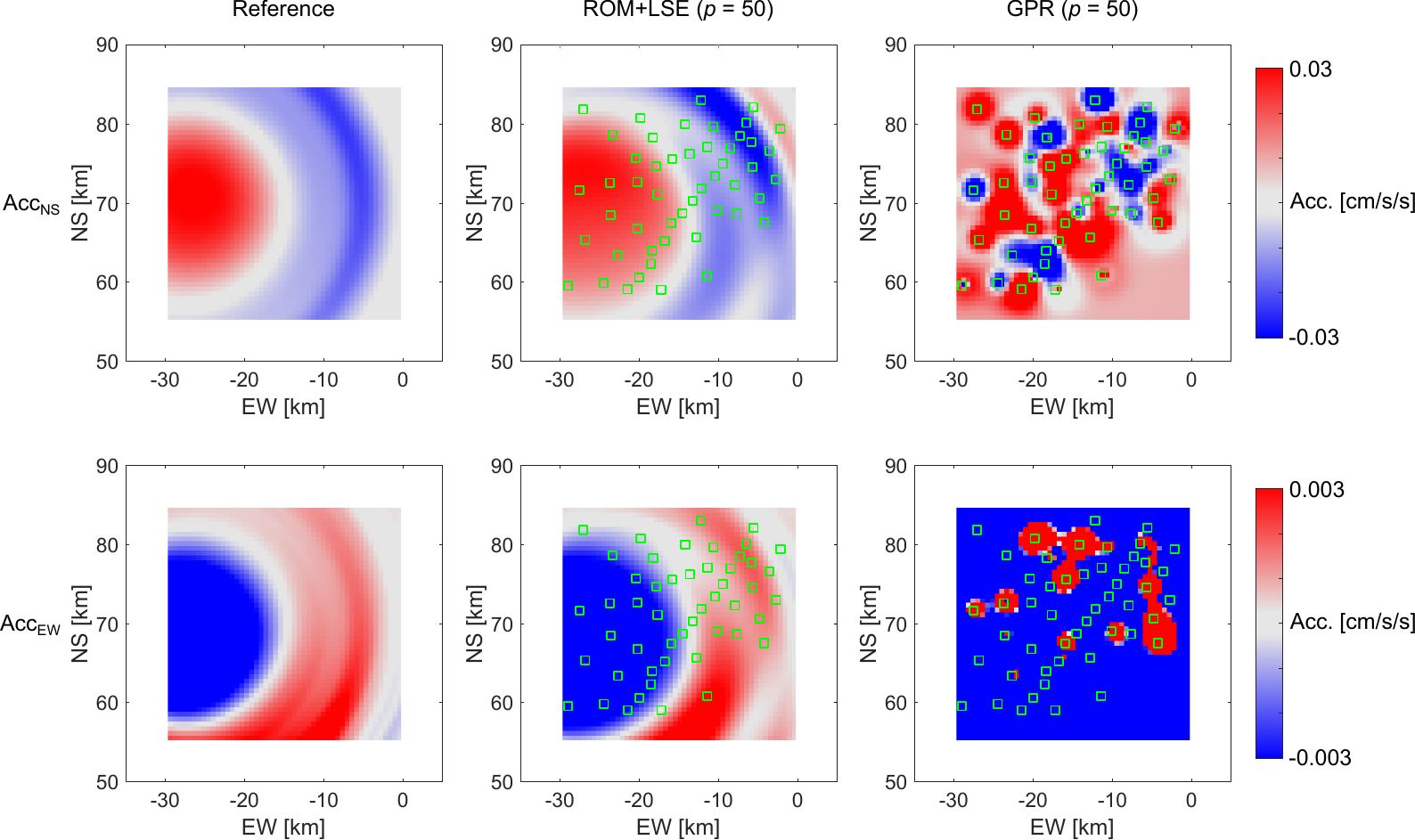}
\caption{Reconstructed wavefields obtained by the ROM-based method and the GPR-based method at $t=24.88$~s in the case of the hypocentral region 2 (w/ observation noise). [This figure appears in colour in the online issue.]}
\label{fig:wavefieldGP_in}
\end{figure*}

In the case of noise-contaminated observations, the quality of the reconstructed wavefield is degraded, as shown in Figures~\ref{fig:wavefieldGP_out} and \ref{fig:wavefieldGP_in}. The GPR-based method obviously failed to reconstruct the wavefields for both cases of the hypocentral regions, even though all available observation sites were used. On the other hand, the ROM-based method almost reproduces the wavefields consistent with the reference data. The ROM-based method utilises the modes of the wavefield as prior information, and thus, the present method is robust to observation noise.

\begin{figure}
\centering
\includegraphics[width=6.5cm]{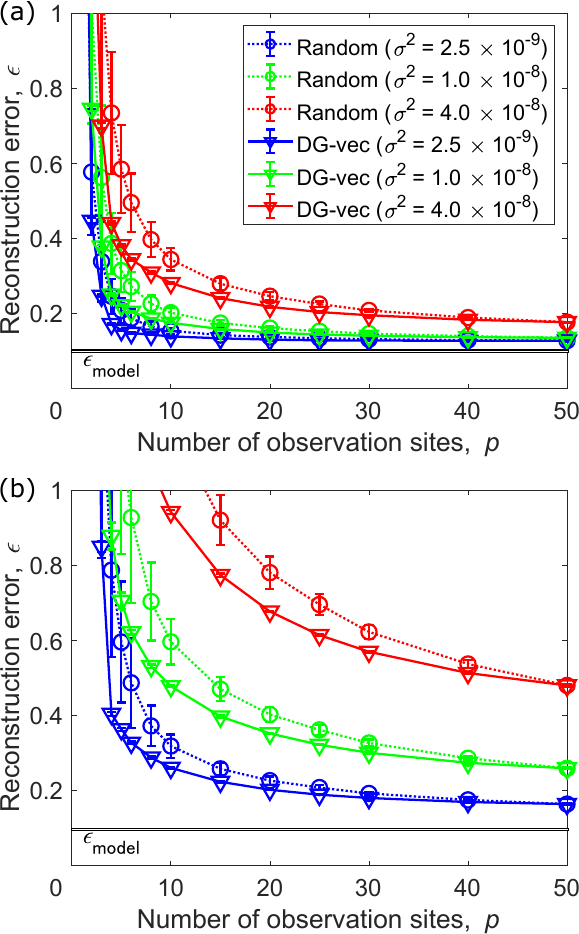}
\caption{\reviewerA{Reconstruction error with respect to the number of observation sites (w/ observation noise) for the method based on ROM and LSE: (a) hypocentral region 1; (b) hypocentral region 2.} [This figure appears in colour in the online issue.]}
\label{fig:reconst}
\end{figure}

The reconstruction error obtained with the noise-contaminated observation is shown in Figure~\ref{fig:reconst}. \reviewerA{In addition to the noise level of $\sigma^2=1.0\times 10^{-4}$, the reconstruction error with lower and higher noise level of $\sigma^2=2.5\times10^{-9}$ and $\sigma^2=4.0\times10^{-8}$ were displayed.} The results of the GPR-based method do not appear in this figure because of the quite larger reconstruction error. The same as in the noise-free condition, the reconstruction error decreases as the number of selected observation sites increases \reviewerA{for all noise levels} and is asymptotic to model error $\epsilon_{\rm model}$. However, the reconstruction error with a large number of observation sites is still larger than that in the case of noise-free conditions due to strong observation noise. 
\reviewerA{The level of the reconstruction error becomes high as the level of the noise becomes high. In addition, a difference in the reconstruction error obtained by observation sites selected by random selection (appear in dotted lines) and by the DG-vec (appear in solid lines) method becomes large for a smaller number of observation sites. The reconstruction error and its variance can be reduced effectively by optimizing the observation sites used for reconstruction.}
\reviewerB{In the present study, the standard deviation of the variation in the hypocentral location is the same for both hypocentral region cases. Therefore, the variation in the wavefield due to the change in the hypocentral location in the case of the hypocentral region 2 is larger than that in the case of the hypocentral region 1 because the distance between the hypocentral region and ROI is short. In addition, the magnitude of acceleration in the case of the hypocentral region 2 is smaller than that in the case of the hypocentral region 1. Therefore, the condition for reconstruction is more difficult in the case of the hypocentral region 2, and the reconstruction error tends to be large compared with that in the case of the hypocentral region 1.}

The decrement of reconstruction error by increasing the number of selected observation sites is large at a smaller number of observation sites. The difference between the reconstruction error obtained by the randomly selected observation sites and the optimized set of observation sites using the DG-vec method becomes small when the number of selected observation sites is large. However, there is a clear difference when the number of selected observation sites is small, and the set of observation sites selected by the DG-vec method shows better performance in terms of the reconstruction error. Since the large observation noise, the standard deviation of the reconstruction error obtained by the randomly selected observation sites is larger than that in the case of the noise-free condition. On the other hand, the standard deviation of the reconstruction error obtained by the observation sites selected by the DG-vec method is kept at the same level as that of the noise-free condition. Hence, the subset of the observation sites selected by the DG-vec method provides a better and more stable estimation of the wavefield, even if the estimation is carried out using a smaller number of observations with large observation noise.


\begin{figure*}
\centering
\includegraphics[width=16.5cm]{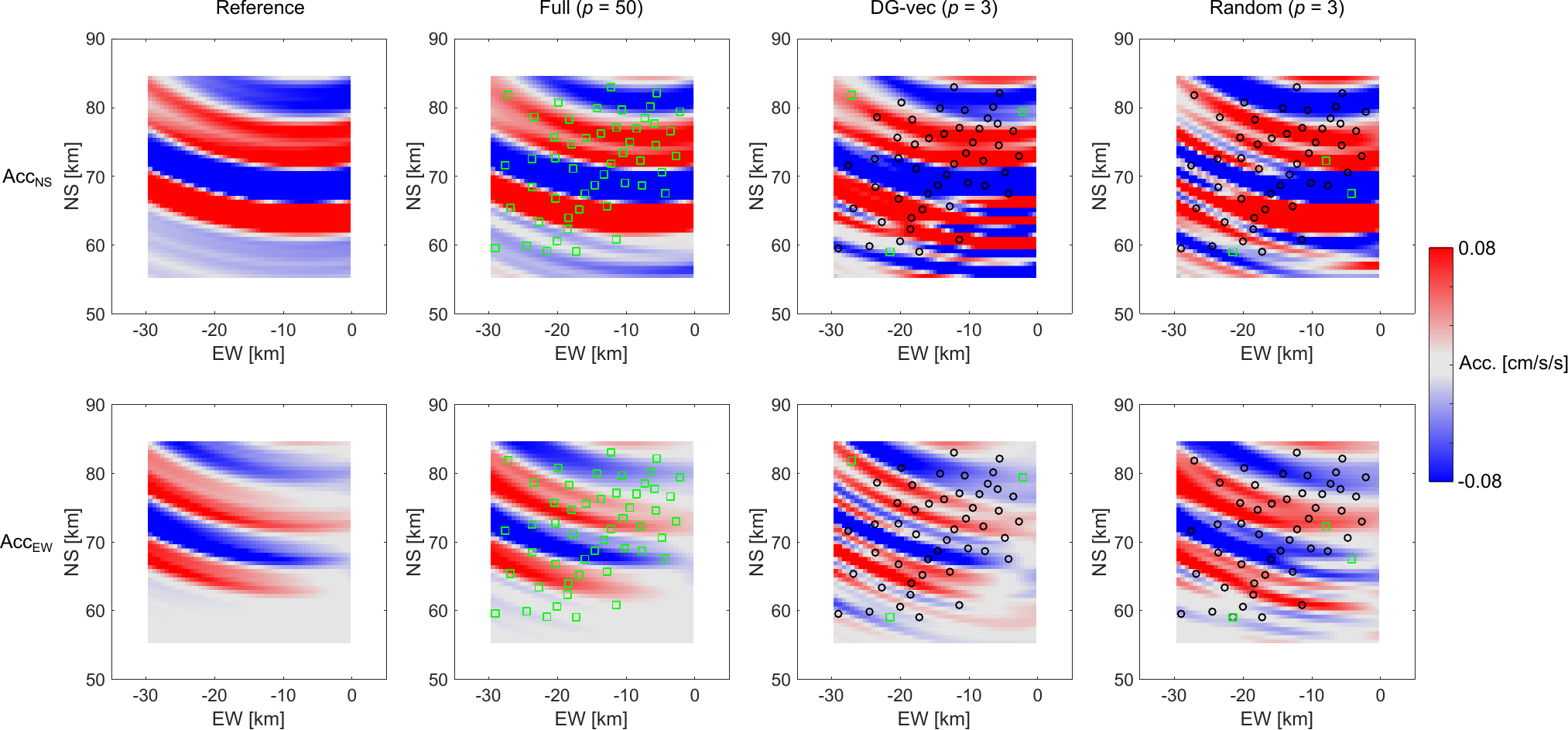}
\caption{Effect of the number of observation sites and selection method on the reconstructed wavefields at $t=24.24$~s in the case of the hypocentral region 1 (w/ observation noise). [This figure appears in colour in the online issue.]}
\label{fig:wavefield_out}
\end{figure*}

\begin{figure*}
\centering
\includegraphics[width=16.5cm]{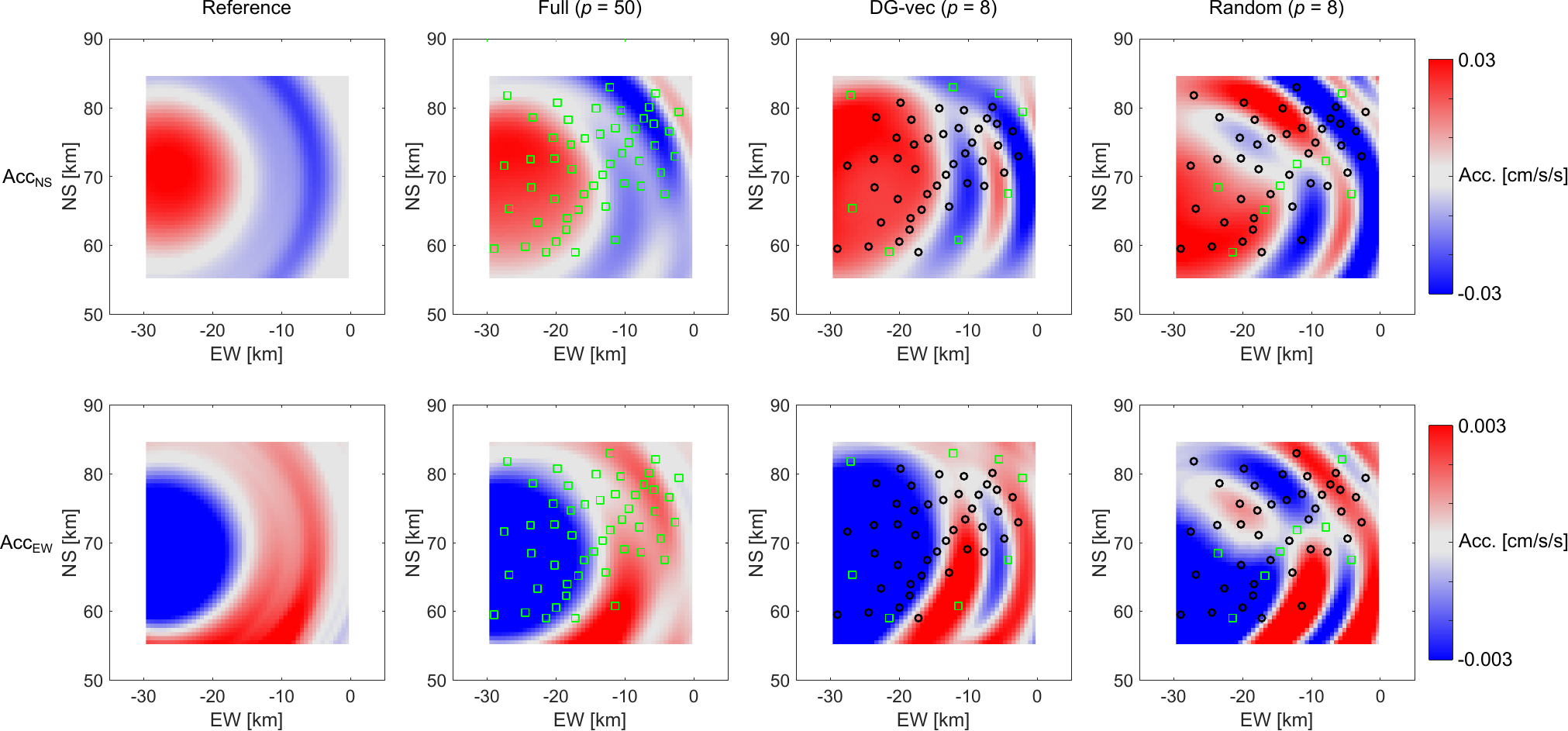}
\caption{Effect of the number of observation sites and selection method on the reconstructed wavefields at $t=24.88$~s in the case of the hypocentral region 2 (w/ observation noise). [This figure appears in colour in the online issue.]}
\label{fig:wavefield_in}
\end{figure*}

Figures~\ref{fig:wavefield_out} and \ref{fig:wavefield_in} show the comparison of the reference and reconstructed wavefields at a certain time. The sets of observation sites were the same as in the case of noise-free observation. The reconstructed wavefield obtained by the limited number of observation sites captures the general characteristics, such as the propagation direction of waves. However, unnatural fluctuations appear in the reconstructed results obtained with the set of the observation sites selected by the DG-vec method and random selection, and the reconstructed wavefield is different from the reference data, unlike the case of noise-free observation. The difference in the results obtained by the observation sites selected by the random selection and the DG-vec method is inconspicuous in the case of the hypocentral region 1. \reviewerB{Hence, the reconstructed wavefield seems to be non-sensitive to the site selection.} On the other hand, the difference can be seen clearer in the case of the hypocentral region 2. In particular, the concentric structure is not kept in the reconstructed wavefield obtained by the randomly selected observation sites. The reconstruction results are much improved when the reconstruction is carried out using all available observation sites. In this case, the reconstructed wavefield is close to the reference data, despite the observation signal being highly contaminated by noise.

\begin{figure}
\centering
\includegraphics[width=12.5cm]{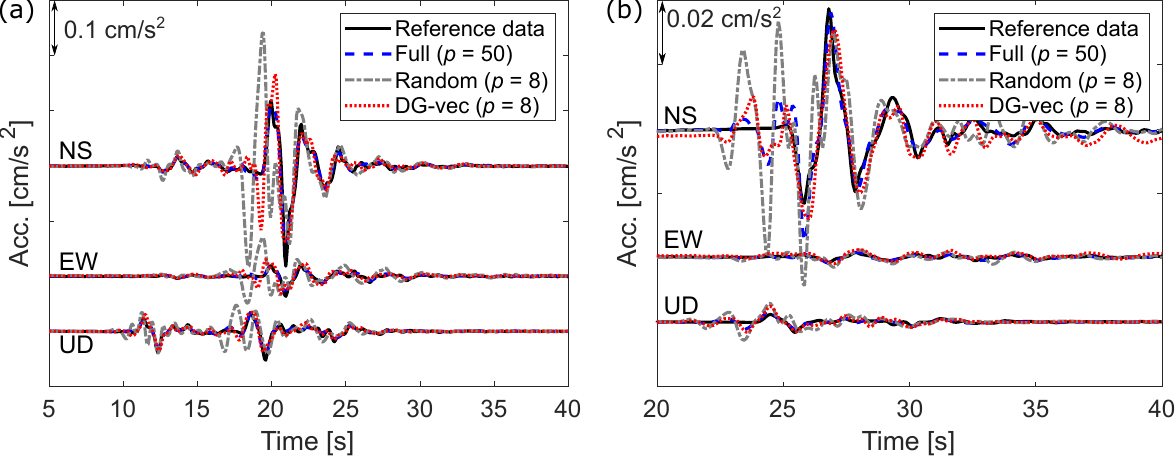}
\caption{\reviewerB{Reconstructed waveform in the center of the ROI (w/ observation noise): (a) hypocentral region 1 and (b) hypocentral region 2.}}
\label{fig:waveform}
\end{figure}

Figure~\ref{fig:waveform} shows the reconstructed waveform. The reconstructed data were degraded due to observation noise, as discussed in the reconstructed wavefield. The reconstructed waveform obtained with $p=50$ shows good agreement with the reference waveform. Since the signal-to-noise ratio of the observation is lower, the difference in the reference waveform and reconstructed waveform obtained with $p=50$ in the case of the hypocentral region 2 is larger than in the case of the hypocentral region 1. The reconstructed waveform becomes more different from that of the reference data with decreasing the number of observation sites. Particularly, the fluctuations with large amplitudes appear immediately before and \reviewerB{early parts} of the S-wave. In the case of the waveform, the difference between the reconstructed data by the set of the observation sites selected by the random selection and the DG-vec method is clearer than in the case of the wavefield. The artificial fluctuations immediately before and \reviewerB{early parts} of the S-wave are effectively reduced by using the set of observation sites selected by the DG-vec method. \reviewerB{This is because the site selection is conducted based on the optimal design of experiment, and the estimation error in the latent variable is minimized in the case of the DG-vec method is used.}

\begin{figure}
\centering
\includegraphics[width=6.5cm]{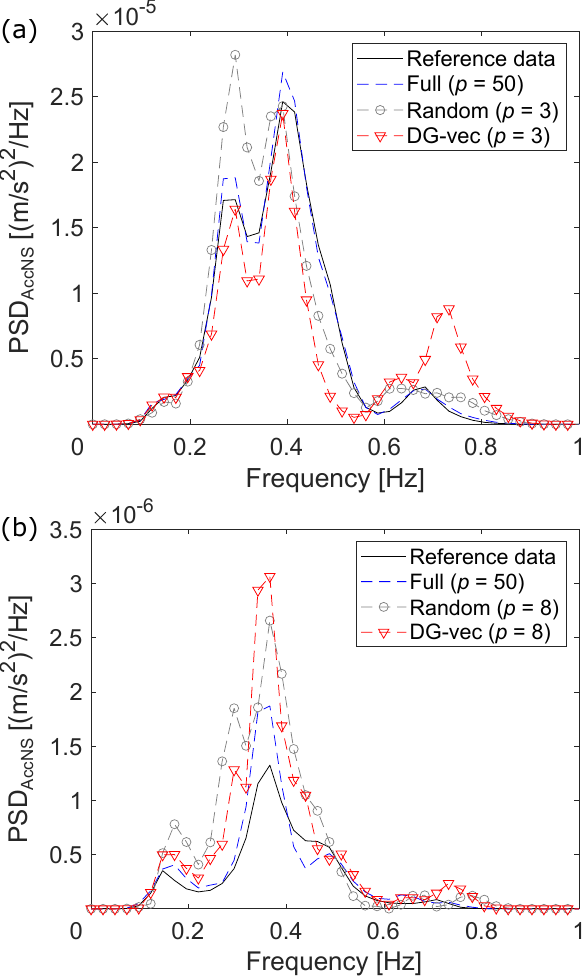}
\caption{Frequency spectra of the reconstructed data at the center of ROI (w/ observation noise): (a) hypocentral region 1 and (b) hypocentral region 2.}
\label{fig:PSDreconst_25-50}
\end{figure}

Figure~\ref{fig:PSDreconst_25-50} shows the comparison of the reconstructed waveforms in the frequency domain. The frequency spectrum of the reconstructed data obtained by $p=50$ observation sites shows good agreement with that of the reference data in the case of the hypocentral region 1. Although the difference in the frequency spectra becomes noticeable by decreasing the number of observation sites used for the seismic wavefield reconstruction, the global nature of the frequency spectrum of the reconstructed data, such as the peak frequency and the number of peaks, is similar to that of the reference data. The variance in the reconstruction error is large when a set of randomly selected observation sites is used. The trend of the influence of the number of observation sites used for the reconstruction is similar to that in the case of the hypocentral region 2. In both cases, the frequency spectra of the reconstructed data obtained by a limited number of observation sites tend to overestimate PSD at the peak frequency, and this trend is emphasized when randomly selected observation sites are used for reconstruction. 

\section{Conclusions}\label{sec:conclusion}
The present study proposed a seismic wavefield reconstruction framework based on compressed sensing using the data-driven ROM. The proposed framework was applied to simulation data, and the wavefield reconstruction was conducted using the noise-free observation and noise-contaminated observation.

For the reconstruction based on the noise-free observation, the reconstruction error is reduced to the model error, even though the number of observation sites used for reconstruction is limited and randomly selected. The reconstructed data reproduce the frequency spectra and time-series waveform of the reference data. In addition, the reconstruction error obtained by the proposed framework is much smaller than that obtained by reconstruction using the GPR-based method.

For the reconstruction based on the noise-contaminated observation, the accuracy of the reconstructed wavefield was degraded due to the observation noise, but the reconstruction error obtained with all available ($p=50$) observation sites is close to the model error. On the other hand, the reconstructed wavefields obtained by the GPR-based method collapse completely, even though all available observation sites were used. Although the reconstruction error is greater than that obtained with $p=50$, the number of observation sites to be used for reconstruction can be reduced while minimizing deterioration and scatter of the reconstructed data by combining with the DG-vec method. The global nature of the frequency spectrum of the reconstructed data, such as the peak frequency and the number of peaks, is similar to that of the reference data. When reducing the number of observation sites, the difference becomes clearly appears in the time-series waveform. In particular, the artificial fluctuations with large amplitude appear immediately before and \reviewerB{early parts} of the S-wave in the reconstructed data with a reduced number of observation sites. The artificial fluctuations immediately before and \reviewerB{early parts} of the S-wave are effectively reduced by using the set of observation sites selected by the DG-vec method.

The subset of the optimized observation sites by the DG-vec method provides a better and more stable estimation of the wavefield, even if the estimation is made with a smaller number of observations with large observation noise. The difference in the performance of the set of observation sites can be minimized by using the optimization method of the observation sites, such as the DG-vec method, and the validation of the reconstructed wavefield can be conducted more properly than that using the sets of randomly selected observation sites.

The framework proposed in this study provides quick and high-precision seismic wavefield reconstruction. Although the present framework requires the construction of the ROM and the selection of the set of observation sites for each epicenter area, these prepossess can be executed offline in advance, and thus, the ROM and the set of observation sites for reconstruction can be determined instantly based on the provisional hypocenter information. Since computational resources are limited, the ROM of the seismic wavefield was constructed using simulations with a 1-D velocity structure in the present study. \reviewerB{Therefore, there are differences between a real wavefield and a synthesized one using a 1-D velocity structure depending on the underground structure. This point is one of the major issues in applying the present framework to real datasets, but the influence of the difference in the velocity structure can be reduced by using the high-fidelity simulation with a 3-D velocity structure. Although the computational cost of the offline analysis for the construction of the ROM becomes high, the influence of the difference in the velocity structure can be reduced by using the high-fidelity simulation with a 3-D velocity structure. In addition, the velocity structure itself can also be included as the parameter when constructing the ROM. However, high-fidelity simulation requires a large computational cost, and the objective of the present paper is to propose a seismic wavefield reconstruction framework based on compressed sensing using data-driven ROM. Thus, the numerical simulation with a 1-D subsurface structure model was employed in the present study. It should be noted that the waveform calculated using a 1-D velocity structure in the low-frequency band (approximately lower than 0.2~Hz) shows good agreement with that calculated using a 3-D velocity structure. Therefore, reconstruction of low-frequency waves, such as direct waves from epicenters with a period longer than 10 seconds that can damage large structures in a metropolitan area, is possible even if the simulation uses a 1-D subsurface structure.}
\reviewerB{In addition, another strategy is to construct a noise model to correct for the influence of noise due to errors in the model. The noise caused by errors in the model might be constructed based on the difference between the observed waveform and the reconstructed waveform at the locations of observation sites.}

\section*{Acknowledgment}
This work was supported by JST CREST (JPMJCR1763).

\section*{Data availability}
The code developed by \citet{hisada1995efficient} for the numerical calculation of the theoretical waveforms is available through the website at \url{http://kouzou.cc.kogakuin.ac.jp/Open/Green/.}
The data of earthquake mechanism information was provided by the National Research Institute for Earth Science and Disaster Resilience at \url{https://www.fnet.bosai.go.jp/event/joho.php?LANG=en.}

\bibliographystyle{gji}
\bibliography{main.bib}




\end{document}